\documentclass{lmcs}
\pdfoutput=1
\usepackage[utf8]{inputenc}

\usepackage{lastpage}
\lmcsdoi{21}{4}{3}
\lmcsheading{}{\pageref{LastPage}}{}{}%
{Jan.~23,~2024}{Oct.~03,~2025}{}

\usepackage[T1]{fontenc}
\usepackage{xcolor}
\usepackage{acronym}
\usepackage{amsmath}
\usepackage{amssymb}
\usepackage{mathtools}
\usepackage{adjustbox}
\usepackage{ifthen}
\usepackage{bmpsize}
\usepackage{tikz}
\usetikzlibrary{arrows,shapes.geometric,positioning,automata,patterns}
\usetikzlibrary{trees,decorations.pathmorphing,calc}
\usetikzlibrary{intersections}
\usepackage{mathtools}
\usepackage{relsize}
\usepackage[inline]{enumitem}
\newlist{inlinelist}{enumerate*}{1}
\setlist[inlinelist]{label=}
\usepackage{tikz}
\usetikzlibrary{arrows,shapes.geometric,positioning,automata,patterns}
\usetikzlibrary{trees,decorations.pathmorphing,calc}
\usetikzlibrary{intersections}
\usetikzlibrary{decorations.pathreplacing,external,backgrounds}
\usepackage{etoolbox} 

\definecolor{ddarkgreen}{rgb}{0,0.5,0}

\tikzset{-,
  ev/.style={circle,draw, inner sep=1.8pt, minimum size=2pt, outer sep=0.5mm},
  point/.style={circle,fill=black, inner sep=1.8pt, minimum size=2pt, outer sep=0.5mm},
  pointp/.style={rectangle,fill=black, inner sep=2pt, minimum size=2pt, outer sep=0.5mm},
  event/.style={circle,fill=red, inner sep=3pt, minimum size=2pt, outer sep=0.5mm},
  past/.style={event, fill=ddarkgreen},
  present/.style={event,fill=white,draw},
  future/.style={event,fill=blue},
  locality/.style={circle,fill=white, inner sep=2pt, minimum size=2pt, outer sep=0.5mm},
  locality1/.style={locality,fill=orange!50},
  locality2/.style={locality,fill=gray!50},
  locality3/.style={locality,fill=blue!40},
  locality4/.style={locality,fill=black!30!green},
  locality5/.style={locality,fill=purple!50},
  d/.style={circle,fill=gray, inner sep=2pt, minimum size=2pt},
  da/.style={circle,draw,label={[left]$1$}},
  db/.style={circle,draw,label={[left]$2$}},
  dc/.style={circle,draw,label={[left]$3$}},
  lbl/.style={font=\footnotesize},
  state/.style={circle,draw},
  statein/.style={circle,draw,line width={2pt}},
  stateout/.style={circle,draw,line width={1pt},fill=black!16},
  leadstov/.style={draw,decorate,decoration={snake, amplitude=1.95mm, post=lineto, post length=2mm, segment length=1cm, pre length=0.4cm},->,>=stealth'},
  leadsto/.style={draw,decorate,decoration={snake, amplitude=0.35mm, post=lineto, post length=2mm, segment length=1.3mm},->,>=stealth'}, 
  leadstopast/.style={leadsto,draw=ddarkgreen},
  leadstopresent/.style={leadsto,draw=black},
  leadstofuture/.style={leadsto,draw=blue},
  frastagliato/.style={draw,decorate,decoration={snake, amplitude=0.45mm, post=lineto, post length=0mm, segment length=2.3mm},->,>=stealth'},
  appgraph/.style={rectangle,draw,minimum size=1.5cm},
  appgraphc/.style={ellipse,draw,minimum size=1.5cm},
  program/.style={rectangle,draw,radius=0.1cm},
  scheduler/.style={rounded rectangle,draw,radius=0.1cm,font=\small},
  evpast/.style={green},
  evpresent/.style={red},
  evfuture/.style={blue},
  evconc/.style={gray},
}

\usepackage{hyperref}
\usepackage{cleveref} 

\acrodef{ict}[ICT]{Information and Communication Technology}
\acrodef{iot}[IoT]{Internet of Things}
\acrodef{ci}[CI]{Collective Intelligence}
\acrodef{cps}[CPS]{Cyber-Physical System}
\acrodef{cas}[CAS]{Collective Adaptive System}
\acrodef{dcp}[DCP]{distributed collective process}
\acrodefplural{dcp}[DCPs]{distributed collective processes}
\acrodef{pid}[PID]{process identifier}
\acrodef{xc}[XC]{eXchange Calculus}
\acrodef{fc}[FC]{Field Calculus}

\sloppypar

\makeatletter
\AtBeginDocument
 {
   \def\ltx@label#1{\cref@label{#1}}
   \def\label@in@display@noarg#1{\cref@old@label@in@display{#1}}
\def\label@in@mmeasure@noarg#1{%
    \begingroup%
      \measuring@false%
      \cref@old@label@in@display{#1}
    \endgroup}%
 } %
\makeatother

\usepackage{amsmath}
\usepackage{amssymb}
\usepackage{listings}
\usepackage{stmaryrd}
\usepackage{mathtools}
\usepackage{soul}
\usepackage{comment}

\usepackage{xspace}

\makeatletter
\newcommand\footnoteref[1]{\protected@xdef\@thefnmark{\ref{#1}}\@footnotemark}
\makeatother

\newcommand{\proc}{\mathit{\pi_{P_i}}}
\newcommand{\oproc}{\mathit{O_{P_i}}}
\newcommand{\sproc}{\mathit{s_{P_i}}}
\newcommand{\neighset}{\ensuremath{\mathit{N}}}
\newcommand{\eprec}{\ensuremath{\mathit{p}}}

\newcommand{\FORGET}[1]{}

\newcommand{\revA}[1]{#1} %
\newcommand{\revB}[1]{#1} 
\newcommand{\arevC}[1]{#1} 
\newcommand{\revD}[1]{#1} 

\newcommand{\extLmcsStart}{}
\newcommand{\extLmcsEnd}{}
\newcommand{\extLmcs}[1]{#1}

\definecolor{qedgray}{rgb}{0.31,0.31,0.33}

\newcommand{\cp}[1]{{\left( #1 \right)}}

\newcommand{\ap}[1]{{\langle #1 \rangle}}
\newcommand{\bp}[1]{{\left\lbrace #1 \right\rbrace}}

\newcommand{\auxcolour}{\color{violet}} 
\newcommand{\kcolour}{\color{blue}} 

\newcommand{\spawnxc}{\ensuremath{\mathit{spawn_{XC}}}}
\newcommand{\spawnfc}{\ensuremath{\mathit{spawn_{FC}}}}

\newcommand{\eqhl}[1]{\colorbox{gray!30}{\ensuremath{#1}}} 

\newcommand{\name}{\tau}
\newcommand{\BNFcce}{{\textbf{::=}}}
\newcommand{\BNFmid}{\;\bigr\rvert\;}

\newcommand{\e}{\mathtt{e}}

\newcommand{\main}{\mathtt{main}}
\newcommand{\emain}{\e_{\main}}

\newcommand{\aname}{\mathtt{a}}
\newcommand{\bname}{\mathtt{b}}
\newcommand{\cname}{\mathtt{c}}

\newcommand{\xname}{\mathtt{x}}
\newcommand{\yname}{\mathtt{y}}

\newcommand{\nname}{\mathtt{n}}
\newcommand{\sname}{\mathtt{s}}

\newcommand{\anyvalue}{\mathtt{v}}
\newcommand{\anyvaluealt}{\mathtt{w}}

\newcommand{\lvalue}{\ell}

\newcommand{\funvalue}{\mathtt{f}}

\newcommand{\truevalue}{{\mathtt{True}}}
\newcommand{\falsevalue}{{\mathtt{False}}}

\newcommand{\nv}{\ensuremath{\underline{\anyvaluealt}}}

\newcommand{\FVname}{\mathsf{FV}}
\newcommand{\FV}[1]{\FVname(#1)}
\newcommand{\FTVname}{\mathsf{TV}}
\newcommand{\FTV}[1]{\FTVname(#1)}

\newcommand{\tsinsrel}{\prec}

\newcommand{\funKname}{{\kcolour\mathtt{fun}}^\name}
\newcommand{\funK}{{\kcolour\mathtt{fun}\;}}
\newcommand{\defK}{{\kcolour\mathtt{def}\;}}

\newcommand{\exchangeK}{{\auxcolour\mathtt{exchange}}}
\newcommand{\spawnK}{{\auxcolour\mathtt{spawn}}}

\newcommand{\retsendK}{{\kcolour\mathtt{retsend}}}
\newcommand{\pExchange}[4]{\exchangeK(#1, (#2) \, \toSymFoK \, (#3, #4))}

\newcommand{\sExchange}[3]{\exchangeK(#1, (#2) \, \toSymFoK \, \retsendK\: #3)}
\newcommand{\ifK}{{\kcolour\mathtt{if}}}
\newcommand{\muxK}{{\auxcolour\mathtt{mux}}}

\newcommand{\elseK}{{\kcolour\,\mathtt{else}\,}}

\newcommand{\foldK}{{\auxcolour\mathtt{nfold}}}

\newcommand{\valK}{{\kcolour\mathtt{val}\;}}

\newcommand{\uidK}{{\auxcolour\mathtt{uid}}}
\newcommand{\updateSelfK}{{\auxcolour\mathtt{updateSelf}}}

\newcommand{\toSymFoK}{\mathrm{\texttt{=>}}}
\newcommand{\toSymK}[1][]{\stackrel{#1}{\mathrm{\texttt{=>}}}}

\newcommand{\selfK}{{\auxcolour\mathtt{self}}}

\newcommand{\minK}{\auxcolour\mathtt{min}}

\newcommand{\type}{\textit{T}}

\newcommand{\ltype}{\textit{A}}

\newcommand{\ltypefun}{\textit{K}}

\newcommand{\typeOf}[2]{#1[#2]}
\newcommand{\ltypeOf}[1]{\typeOf{\ltypefun}{#1}}

\newcommand{\tbool}{\mathtt{bool}}

\newcommand{\tpair}{\mathtt{pair}}

\newcommand{\tset}{\mathtt{set}}
\newcommand{\tmap}{\mathtt{map}}

\newcommand{\tfieldOf}[1]{\underbar{\ensuremath{#1}}}

\newcommand{\pairK}{{\auxcolour\mathtt{pair}}}

\newcommand{\tvar}{\alpha}
\newcommand{\typescheme}{\textit{TS}}

\newcommand{\TtypEnv}{\mathcal{A}}

\newcommand{\expTypJud}[3]{#1 \vdash #2 : #3}
\newcommand{\surfaceTyping}[3]{
  \begin{array}{@{\!\,}l@{\;}c@{\!\,}}
    \stackrel{~}{{\tiny \textrm{[#1]}}} & #2 \\ \hline
    \multicolumn{2}{c}{#3}
  \end{array}
}
\newcommand{\nullsurfaceTyping}[2]{
  \surfaceTyping{#1}{}{#2}
}

\newcommand{\deviceS}{\Delta}

\newcommand{\Trees}{\Theta}
\newcommand{\emptyseq}{\bullet}

\newcommand{\senstate}{\sigma}

\newcommand{\envmap}[2]{#1\mapsto #2}

\newcommand{\proj}[2]{{#1} \, |_{#2}}
\newcommand{\evalue}[3]{#1[\envmap{#2}{#3}]}

\newcommand{\ruleNameSize}[1]{{\scriptsize #1}}

\newcommand{\domof}[1]{\textbf{dom}(#1)}

\newcommand{\vtree}{\theta}
\newcommand{\mkvt}[2]{#1 \langle #2 \rangle}
\newcommand{\piB}[1]{\pi^{#1}}
\newcommand{\piBof}[2]{\piB{#1}(#2)}
\newcommand{\piI}[1]{\pi_{#1}}
\newcommand{\piIof}[2]{\piI{#1}(#2)}

\newcommand{\bsopsem}[6]{#1;#3;#2\vdash #4\Downarrow #5; #6}
\newcommand{\eqhlauxopsem}[6]{#1;#3;#2\vdash #4\Downarrow^\ast #5; #6}
\newcommand{\eqhlbsopsem}[6]{#1;#3;#2\vdash #4\Downarrow #5; #6}

\newcommand{\deviceId}{\delta}

\newcommand{\substitution}[2]{#1:=#2}
\newcommand{\applySubstitution}[2]{#1[#2]}
\newcommand{\nameOf}{\textit{name}}

\newcommand{\skiptransition}{\\~\\[-10pt]}

\newcommand{\TreeSet}{\Omega}

\newcommand{\aEventS}[0]{\mathbb{E}}

\newcommand{\eventS}[0]{E}
\newcommand{\eventId}[0]{\epsilon}

\newcommand{\setVS}[0]{\mathbf{V}}

\newcommand{\neigh}{\rightsquigarrow}

\newcommand{\devof}{d}
\newcommand{\sensof}{s}

\newcommand{\dvalue}[0]{\mathrm{\Phi}}

\newcommand{\formula}[0]{\phi}
\newcommand{\formulalt}[0]{\psi}
\renewcommand{\prop}[0]{q}

\DeclareMathOperator{\DX}{\mathrm{Y}}
\DeclareMathOperator{\AX}{\mathrm{AY}}
\DeclareMathOperator{\EX}{\mathrm{EY}}
\DeclareMathOperator{\DU}{\mathrm{S}}
\DeclareMathOperator{\AU}{\mathrm{AS}}
\DeclareMathOperator{\EU}{\mathrm{ES}}
\DeclareMathOperator{\DF}{\mathrm{P}}
\DeclareMathOperator{\AF}{\mathrm{AP}}
\DeclareMathOperator{\EF}{\mathrm{EP}}
\DeclareMathOperator{\EP}{\mathrm{EP}}
\DeclareMathOperator{\DG}{\mathrm{H}}
\DeclareMathOperator{\AG}{\mathrm{AH}}
\DeclareMathOperator{\EG}{\mathrm{EH}}

\DeclareMathOperator{\SI}{\Box} 
\DeclareMathOperator{\SC}{\Diamond} 
\DeclareMathOperator{\SB}{\partial} 
\DeclareMathOperator{\SBI}{\partial^-} 
\DeclareMathOperator{\SBC}{\partial^+} 
\DeclareMathOperator{\SR}{\mathcal{R}} 
\DeclareMathOperator{\ST}{\mathcal{T}} 
\DeclareMathOperator{\SU}{\mathcal{U}} 
\DeclareMathOperator{\SG}{\mathcal{G}} 
\DeclareMathOperator{\SF}{\mathcal{F}} 

\definecolor{dark-gray}{gray}{0}

\renewcommand{\ldots}{%
\mathinner{\ldotp\mkern-4mu\ldotp\mkern-4mu\ldotp}%
}

\lstdefinelanguage{afc2}{
	basicstyle=\normalsize\ttfamily\lst@ifdisplaystyle\footnotesize\fi,
	frame=single,
	basewidth=0.50em,
	sensitive=true,
	morestring=[b]",
	morecomment=[l]{//},
	morecomment=[n]{/*}{*/},
	commentstyle=\color{darkgreen},
	keywordstyle=\color{blue}\textbf, keywords={def,if,else,val,return,send,retsend},
	keywordstyle=[3]\color{violet}, keywords=[3]{updateSelf,self,fst,snd,and,or,nbrDist,uid,mux,pair,distanceTo,broadcast,channelBroadcast,myApp},
	keywordstyle=[4]\color{red}, keywords=[4]{let,in,exchange,nfold}
}

\lstdefinelanguage{xc}{
	basicstyle=\normalsize\ttfamily\lst@ifdisplaystyle\footnotesize\fi,
	frame=single,
	basewidth=0.50em,
	sensitive=true,
	morestring=[b]",
	morecomment=[l]{//},
	morecomment=[n]{/*}{*/},
	commentstyle=\color{darkgreen},
	keywordstyle=\color{blue}\textbf, keywords={def,fun,if,else,val,return,send,retsend},
	keywordstyle=[3]\color{violet}, keywords=[3]{updateSelf,self,fst,snd,and,or,nbrDist,minHood,uid,nfold,min,mux,pair,exchange,spawn,nbr,set},
	keywordstyle=[4]\color{red}, keywords=[4]{let,in}
}

\lstdefinelanguage{fcpp}{
	basicstyle=\footnotesize\ttfamily\lst@ifdisplaystyle\footnotesize\fi,
	frame=single,
	basewidth=0.5em,
	sensitive=true,
	morestring=[b]",
	morecomment=[l]{//},
	morecomment=[n]{/*}{*/},
	commentstyle=\color{darkgreen},
	keywordstyle=\color{blue}, keywords={return,auto,FUN,GEN,ARGS,CODE,CALL,for,if,else,const},
	keywordstyle=[2]\color{violet}\textbf, keywords=[2]{bool,int,double,field,nvalue,unordered_map,device_t,size_t,times_t,real_t,task_t,option,set,vector},
	keywordstyle=[3]\color{orange}, keywords=[3]{fold_hood,min_hood,mux,min,max,INF,or,and,not,mod_self,mod_other,map_hood},
	keywordstyle=[4]\color{red}, keywords=[4]{exchange,nbr,rep,share,spawn}
}

\newcommand{\sysname}{{\textsf{XC}}\xspace}

\newcommand{\excvalue}[0]{nvalue}
\newcommand{\excvalues}{\excvalue{}s}

\definecolor{darkgreen}{rgb}{0,0.5,0}

\lstdefinelanguage{scafi}{
  basicstyle=\normalsize\ttfamily\lst@ifdisplaystyle\footnotesize\fi,
	frame=single,
	escapechar=\%,
  keywords={abstract,case,catch,class,def,%
    do,else,extends,final,finally,%
    for,if,implicit,import,match,mixin,%
    new,null,object,override,package,%
    private,protected,requires,return,sealed,%
    super,this,throw,trait,try,lazy,%
    type,val,var,while,with,yield,forSome},
  otherkeywords={=>,<-,<\%,<:,>:,\#},
  keywordstyle=\color{red}\textbf,
  keywordstyle=[2]\color{blue},
  keywords=[2]{exchange,branch,@@},
  keywordstyle=[3]\color{violet},
  keywords=[3]{fold,mux,collect,pair,broadcast,foldSum,distanceTo,channel,timestamp,channelBroadcast},
  keywordstyle=[4]\color{Emerald},
  keywordstyle=[5]\color{Brown},
  sensitive=true,
  morecomment=[l]{//},
  morecomment=[n]{/*}{*/},
  commentstyle=\color{darkgreen},
  morestring=[b]",
  morestring=[b]',
  morestring=[b]"""
}

\newcommand{\supplier}[0]{neighbour}


\begin{document}
\title[Programming Distributed Collective Processes in XC]{Programming Distributed Collective Processes\texorpdfstring{\\}{}
in the eXchange Calculus
}
%
%

\author[G.~Audrito]{Giorgio Audrito \lmcsorcid{0000-0002-2319-0375}}[a]	

\author[R.~Casadei]{Roberto Casadei\lmcsorcid{0000-0001-9149-949X}}[b]	

\author[F.~Damiani]{Ferruccio Damiani\lmcsorcid{0000-0001-8109-1706}}[a]

\author[G.~Torta]{Gianluca Torta\lmcsorcid{0000-0002-4276-7213}}[a]	

\author[M.~Viroli]{Mirko Viroli\lmcsorcid{0000-0003-2702-5702}}[b]	

\address{Universit\`{a} di Torino, Turin, Italy}	
\email{giorgio.audrito@unito.it,
ferruccio.damiani@unito.it,
gianluca.torta@unito.it}  

\address{Alma Mater Studiorum---Università di Bologna, 
Cesena, Italy}	
\email{roby.casadei@unibo.it, mirko.viroli@unibo.it}  

\begin{abstract}
Recent trends like the \ac{iot}
 suggest a vision of dense and multi-scale deployments
 of computing devices in nearly all kinds of environments.
A prominent engineering challenge
 revolves around programming the collective adaptive behaviour
 of such computational ecosystems.
This requires abstractions
 able to capture concepts like
 ensembles (dynamic groups of cooperating devices)
 and collective tasks (joint activities carried out by ensembles).
%
In this work,
 we consider collections of devices interacting with neighbours and that execute in nearly-synchronised sense--compute--interact rounds,
 where the computation is given by a \extLmcs{single program mapping sensing values and incoming messages to output and outcoming messages.}
To support programming whole computational collectives, we propose the abstraction
 of a distributed collective process, 
 which can be used to define at once the ensemble formation logic and its collective task.
We formalise the abstraction in the eXchange Calculus (XC),
 a core functional language based on neighbouring values
 (maps from neighbours to values)
 where state 
 and interaction
 is handled through a single primitive, \texttt{exchange}\extLmcs{,
 and provide a corresponding implementation in the FCPP language.
Then, we exercise 
 distributed collective processes
 using two case studies: multi-hop message propagation and
 distributed monitoring of spatial properties.
Finally,
} we discuss the features of the abstraction
 and
 its suitability
 for different kinds of
 distributed computing applications.
\end{abstract}

\maketitle              

\def\figResizeFactor{1}

\section{Introduction}\label{sec:intro}

Programming the collective behaviour
 of large groups of interactive computing devices
 is a major research problem,
 urged by scenarios like
 the Internet of Things~\cite{DBLP:journals/cn/AtzoriIM10}
 and swarm robotics~\cite{DBLP:journals/swarm/BrambillaFBD13}.
This problem is 
 (partially) addressed
 by several related research threads
 including
 \emph{coordination}~\cite{DBLP:journals/cacm/GelernterC92,DBLP:journals/ac/PapadopoulosA98},
  \emph{multi-agent systems}~\cite{boissier2020maop},
 \emph{collective adaptive systems}~\cite{DBLP:journals/sttt/NicolaJW20},
 \emph{macroprogramming}~\cite{Casadei2023,regiment}, and
 \emph{spatial computing}~\cite{SpatialIGI2013}.

This activity can be supported by
 suitable \emph{programming abstractions}
 supporting declarative specifications
 of collective behaviours.
Examples of abstractions include
 ensembles~\cite{DBLP:journals/taas/NicolaLPT14},
 computational fields~\cite{vbdacp:survey},
 collective communication interfaces~\cite{DBLP:conf/nsdi/WelshM04,abd2020programming-cas-attribute-based},
 and collective-based tasks~\cite{DBLP:journals/tetc/ScekicSVRTMD20}.
In this work,
 we cover the abstraction of a \emph{\ac{dcp}},
 inspired by aggregate processes~\cite{casadei2019aggregate,casadei2021engineering,DBLP:journals/percom/TestaADT22},
\extLmcs{ which provides a way to specify
  collective activities
 that run concurrently and spread out on evolving domains of devices.

\revB{In this paper,} we interpret the \ac{dcp} mechanism
 in the abstract model of event structures~\cite{nielsen:event_structures,abdv:universality},
 and present its realisation in terms of a programming construct in the \ac{xc}~\cite{DBLP:conf/ecoop/AudritoCDSV22}.
\ac{xc} is a core functional language that generalises field calculi~\cite{vbdacp:survey} and promotes  programming the self-organisation in systems of neighbour-interacting devices.
The \ac{xc}-based formalisation of \acp{dcp}
 is also implemented in the FCPP language~\cite{a:fcpp}.
Then, we discuss the abstraction,
 and present examples of use of \acp{dcp},
 showing they can capture several patterns of
 collective adaptive behaviour and self-organisation.
We use the FCPP simulator~\cite{DBLP:conf/coordination/AudritoRT22} to assess that \ac{dcp}-based \ac{xc} programs provide the intended functionality.

\revB{
In summary,
  we build on previous work on the field calculus and aggregate processes~\cite{casadei2021engineering} (see \Cref{sec:rw} for details),
  and provide the following novel contributions:
\begin{itemize}
\item we propose the \ac{dcp} abstraction as a generalisation of
 the field calculus-based mechanism of aggregate processes
  where process propagation
  can be controlled by the sender (and not just through the \emph{opt-out} of the recipient);
\item we formalise and implement the \ac{dcp} abstraction in the \ac{xc} formal language and the FCPP programming language;

\item we use simulated case studies (publicly available at a GitHub repository---cf. \Cref{footnote:eval-repo}) and examples to discuss the features of the \ac{dcp} abstraction
 and show performance benefits w.r.t. aggregate processes.
\end{itemize}
}

This manuscript is an extended version of  paper~\cite{DBLP:conf/coordination/AudritoCDTV23} presented at COORDINATION'23.
Beyond such content,
 it includes a self-contained formalisation (\Cref{sec:background-xc,sec:contrib}),
 a description of the implementation in the FCPP language (\Cref{sec:fcppimpl}),
 a whole new set of simulation-based experiments (\Cref{eval:msg-propagation,sec:eval:pastctl}),
 additional examples and discussion (\Cref{sec:disc}),
 and a larger coverage of related work (\Cref{sec:rw}).
}

\extLmcs{
The remainder of the paper is structured as follows.
\Cref{sec:motiv} positions the work, explicating its context and motivation.
\Cref{sec:background-xc} provides background on XC.
\Cref{sec:contrib} provides the contribution: formalisation, description of the abstraction, and implementation.
\Cref{sec:eval} presents two \revB{simulated use cases} that exploit the proposed techniques\revB{, where
  quantitative metrics are extracted to assess correctness and resource usage}.
\Cref{sec:disc} \revB{provides a broader discussion of the proposed programming framework,
  together with further examples
  to highlight its features
  and its applicability}.
\Cref{sec:rw} covers related work.
\Cref{sec:conc} provides a wrap-up \revB{with further insights on limitations to be addressed in future work}.
}

\section{Context and Motivation}\label{sec:motiv}

This work lies in the context of models and languages for programming collective behaviour~\cite{DBLP:journals/sttt/NicolaJW20,casadei2023artl-ci,DBLP:journals/swarm/BrambillaFBD13}.
Indeed, achieving the desired collective behaviour is an engineering goal for different domains and applications:
\begin{itemize}
\item \emph{Swarm robotics}. Multiple robots may be tasked to move and act as a collective to explore an unknown environment~\cite{DBLP:journals/scirobotics/McGuireWTKC19}, to search and rescue victims for humanitarian aid after disasters~\cite{DBLP:conf/cogsima/ArnoldJAM20}, to map a crop field for the presence of weeds~\cite{DBLP:conf/avss/AlbaniIHT17}, to transport objects \extLmcs{whose weight exceeds the limit of individual robots}~\cite{DBLP:journals/ijbic/GrossD09}, etc.
\item \emph{The \ac{iot}}. The \emph{things} should coordinate to promote application-level goals (e.g., by gathering and processing relevant data) while making efficient use of resources. For instance,
  the nodes could
  support the aggregation of machine learning models~\cite{DBLP:conf/bigdataconf/SudharsanYNKB21},
  or collaborate to
  measure the collective status of the network to support various activities ranging from environment sensing~\cite{DBLP:conf/percom/LiuH019} to remote attestation of system integrity~\cite{DBLP:journals/comsur/AmbrosinCLRR20}.
\item \emph{Hybrid \ac{ci}}. Socio-technical systems involving humans and computing devices could be programmed as ``social machines''~\cite{DBLP:journals/ai/HendlerB10}
  executing orchestrated tasks~\cite{DBLP:journals/tetc/ScekicSVRTMD20},
  or fostering the emergence of collective knowledge~\cite{DBLP:journals/ws/Gruber08}.

\item \emph{Computing ecosystems}. Modern infrastructures spanning the edge-cloud continuum can be conceived as collective systems. The computing nodes could share and process information to create efficient topologies~\cite{DBLP:journals/percom/KaragiannisS21,DBLP:journals/fgcs/PianiniCVN21}
 to coordinate task allocation and execution~\cite{DBLP:conf/ciot/MohanK16}, resiliently.
\end{itemize}
This problem is at the core of
 several related research threads.
The field of \emph{coordination} ~\cite{DBLP:journals/cacm/GelernterC92,DBLP:journals/ac/PapadopoulosA98} addresses it by governing interaction;
 \emph{collective adaptive systems} engineering~\cite{DBLP:journals/sttt/NicolaJW20}  investigates means for collective adaptation in large populations of agents;
 \emph{spatial computing}~\cite{SpatialIGI2013}
 leverages spatial abstractions to drive behaviour and perform computation about and \emph{in} space;
 \emph{macroprogramming} ~\cite{Casadei2023,regiment} takes a programming language-perspective to expressing and steering macroscopic behaviour;
 \emph{multi-agent systems}~\cite{boissier2020maop} consider how autonomous entities can solve problems through cooperation, competition, cognitive mechanisms, organisational structures, etc.

In this work, we consider a language-based software engineering perspective~\cite{DBLP:journals/scp/Gupta15}.
\extLmcs{In the literature, several abstractions have been proposed: ensembles, collective communication interfaces, spatiotemporal regions, etc. (see \Cref{sec:rw} for details).
Similarly, we seek for abstractions for expressing collective behaviour, with some requirements and reference contexts that we delineate in the following.
}

In particular, we consider \emph{collective systems},
 namely largely homogeneous collections
 of devices or agents.
Each device can be thought of as a resource
 that provides capabilities
 and provides access to a local context
 that depends on its situation on the environment
 and possibly its state.
For instance, in a smart city, fixed smart lights may be located nearby streets, smart cameras may support monitoring of facilities, smart vehicles may move around to gather city-wide infrastructural data, etc.
To reduce bottlenecks and single-points-of-failure, we avoid centralisations and opt for a fully decentralised approach: a device may interact only within its local context (e.g., a portion of the environment
 and other nearby devices).
If our goal is to exploit the distributed, pervasive computer
 made of an entire collection of situated devices,
 an idea could be to run \emph{collaborative} tasks
 involving subsets of devices---to exploit their resources, capabilities, and contexts.
Since a process may not know beforehand
 the resources/capabilities it needs
 and the relevant contexts,
 it may embed the logic to look for them,
 i.e., to spread over the collective system
 until its proper set of supporting devices
 have been identified.
Moreover, the requirements of the process may change over time,
 dynamically self-adapting to various environmental conditions
 and changing goals.
Within a process that concurrently spans a collection of devices, local computations may be scheduled
and information may flow around in order to support collective activities~\cite{DBLP:journals/swarm/BrambillaFBD13,Wood2009} such as collective perception~\cite{DBLP:conf/vnc/GuntherRWF16}, collective decision-making~\cite{Bulling2014ma-decison-making}, collective movement~\cite{Navarro2013coll-mov}, etc.
So, if the collective that sustains the process
 decides that more resources are needed,
 the process may spread to a larger set of devices;
 conversely, if the collective task has been completed,
 the devices may quit the process,
 eventually making it vanish.
Informally, this is our idea of a \emph{\aclp{dcp} (\acs{dcp})}:
 a \emph{process} (i.e., a running program or task)
 which is \emph{collective} (i.e., a joint activity carried out by groups of devices)
 and \emph{distributed} (i.e., concurrently active on multiple devices),
 where the collective task and the underlying ensemble
 can mutually affect each other,
 and ensemble formation is driven by decentralised device-to-device interaction.

\revB{
\paragraph{Approach Overview}
In this paper, we aim to characterise the aforementioned abstract idea of a \ac{dcp}
 in terms of the following main ingredients:
\begin{itemize}
\item \emph{event structures}: this formal model will serve to represent collective system executions (i.e., networks of device computation rounds causally-related by message-passing)
 and to describe \acp{dcp} in a general setting;

\item \emph{\acf{xc}}: this core language
  captures a minimal set of constructs
  for expressing collective computations
  (i.e., the activity to be performed at the different events of a collective system execution); we will extend the \ac{xc} language to provide a \emph{programming interface for \acp{dcp}}.
\end{itemize}
The following section provides background on these two ingredients.
}

\section{Background: the eXchange Calculus}
\label{sec:background-xc}

We consider the \emph{eXchange Calculus (XC)}~\cite{DBLP:conf/ecoop/AudritoCDSV22}
 as the formal framework
 for
 modelling,
 reasoning about,
 and implementing \acp{dcp}.
In this section, \revB{we start by providing an overview of the calculus (\Cref{sec:background-xc-overview}), then}
present the system and execution model
 (\Cref{sys-model}),
 providing an operational view of the kinds of systems we target,
 and later
 describe the basic constructs of XC
 that we leverage in this work (\Cref{ssec:data}).
 \revB{Finally, we conclude with the formal details of \sysname: syntax (\Cref{ssec:pfc:syntax}), typing (\Cref{ssec:typing}), and semantics (\Cref{ssec:opsem}).}

\revB{
\subsection{Overview}\label{sec:background-xc-overview}
XC is a core functional language capturing a minimal set of constructs for programming the self-organising behaviour of collections of neighbour-interacting devices (as in robot swarms, sensor networks, and \ac{iot} systems).
It assumes a system model
 where a collection of devices (equipped with sensors and actuators for interacting with their environment)
 operate in asynchronous \emph{sense--compute--interact} rounds.
An XC program expresses the logic at the ``compute'' step:
 it receives as input the local context of the device (i.e., sensor values and data from neighbours)
 and returns as output the actuations to perform
 and the data to be shared with neighbours.

Structurally, XC is basically a lambda calculus
 extended with the \emph{neighbouring value (nvalue)} data type
 and a single communication construct called \emph{exchange}.
XC generalises over field calculi~\cite{vbdacp:survey,DBLP:journals/lmcs/AudritoCDV23}, which historically had \emph{(i)}
 no support for sending different messages to different neighbours,
  and \emph{(ii)} different constructs for state management and communication.
An nvalue is a map $\underline{\anyvaluealt}$ from device identifiers $\deviceId_i$ to local values $\lvalue_i$, with an additional \emph{default} value $\lvalue$ for missing keys, written as $\underline{\anyvaluealt} = \lvalue[\deviceId_1 \mapsto \lvalue_1, \ldots, \deviceId_n \mapsto \lvalue_n]$.
Construct $\pExchange{\e_i}{\underline{\nname}}{\e_r}{\e_s}$ evolves a distributed quantity in space and time by mapping the nvalue of neighbour messages $\underline{\nname}$ (using the value from $\e_i$ as default) to a return value (computed in expression $\e_r$) and a value to be sent (expression $\e_s$).

To formalise XC,
 it is typical (as for other field calculi)
 to distinguish the overall system execution (i.e., environment evolution, exchange of messages, scheduling of rounds)
 from the execution of individual rounds.
Accordingly, XC is formalised in terms of
  \emph{(i)} an operational semantics describing how a device interprets an XC program,
  and
  \emph{(ii)} a network semantics expressing the relationships between rounds, in the framework of \emph{event structures}.

XC can succinctly encode basic self-organisation building blocks, which can in turn be composed to implement more complex collective behaviours.
For instance, a gradient, namely the program computing distance estimates from each device to their closest source device, can be re-used to provide optimal directions for propagating and collecting distributed information.
The ability to program emergent behaviour
 based on repeated execution of
 programs supporting the processing and diffusion of local information
 is especially instrumental to build \acp{dcp}---as it will be covered in \Cref{sec:contrib}.
}

\definecolor{viridisYellow}{RGB}{253,231,37}
\definecolor{viridisGreen}{RGB}{94,201,98}
\definecolor{viridisDarkgreen}{RGB}{33,145,140}
\definecolor{viridisBlue}{RGB}{59,82,139}
\definecolor{viridisPurple}{RGB}{68,1,84}

\newcommand{\colorFuture}{viridisBlue!30}
\newcommand{\colorPresent}{viridisYellow!60}
\newcommand{\colorPast}{viridisGreen!30}

\newcommand{\viridisBlue}{viridisBlue!40}
\newcommand{\viridisGreen}{viridisGreen!30}

\newcommand{\colorRegionA}{viridisBlue!40}
\newcommand{\colorRegionB}{viridisYellow!60}
\newcommand{\colorRegionC}{viridisGreen!40}
\newcommand{\colorRegionOpacity}{0.6}

\newcommand{\esResizeFactor}{1.0}

\subsection{System Model}
\label{sys-model}

The target system that we would like to program
 can be modelled as a collection of \emph{nodes},
 able to interact with the environment through \emph{sensors} and \emph{actuators}, and
 able to communicate with \emph{neighbours} by exchanging messages.
We assume that each node runs in asynchronous \emph{sense--compute--act} \emph{rounds}, where
\begin{enumerate}
\item \emph{sense}: the node queries sensors for getting up-to-date environmental values, and gathers recent messages from neighbours (which may expire after some time)---all this information is what we call as the node's \emph{context};
\item \emph{compute}: the node evaluates the common control program, mapping the context (i.e., inputs from sensors and neighbours) to an output describing the actions to perform (i.e., actuations and communications);
\item \emph{act}: the node executes the actions as dictated by the program, possibly resulting into environment change or message delivery to neighbours.
\end{enumerate}
This kind of loop is used to ensure that the context is continuously assessed (at discrete times),
and the reactions are also computed and performed continuously.
This model has shown to be particularly useful to specify self-organising and collective adaptive behaviours, especially for long-running coordination tasks~\cite{vbdacp:survey}.
\extLmcs{
Note that details regarding aspects like the topology (as defined by a neighbouring relationship), round scheduling, message expiration, and how actual communication is carried out
can be defined by middleware implementations and also tuned on a per-application basis.
What is key here is the ``repeated'' evaluation and action on the local context by each device; upon such change acquired by devices, computations regulate how the novel information is integrated with past information, and how all this is transformed and shared with neighbours.
Flowing from neighbourhood to neighbourhood, the effects of computation would \emph{eventually spread and stabilise~\cite{viroli:selfstabilisation} on a global scale} (cf. \emph{emergence}~\cite{DBLP:conf/saso/WolfH07}).
So, in this work, our focus is on computations that are long-running, ``progressive'', and collective (based on collaboration of groups of devices).
}

\begin{figure}
\newcommand{\toAbove}{1.5cm}
\newcommand{\toRight}{1cm}
\centering
\adjustbox{max width=\esResizeFactor\columnwidth}{
\begin{tikzpicture}
\node[] (origin) at (0,0) {};

\foreach \dev/\evs/\offx/\freq in {%
1/3/ -0.8 / 3.5,%
2/5/ -0.2 / 2.3,%
3/4/ 0.3 / 2.6,%
4/6/ -0.2 / 1.8,%
5/3/ -0.8 / 3.8}
  \foreach \ev in {1,...,\evs}
        \node[present,font=\Large] (d\dev\ev) [above right = \dev*\toAbove and \ev*\toRight*\freq,xshift=-1cm+\offx*2cm,yshift=-1cm] {$\eventId_\ev^\dev$}; 

\makeatletter
\long\def\ifnodedefined#1#2#3{%
    \@ifundefined{pgf@sh@ns@#1}{#3}{#2}%
}
\makeatother
\foreach \ref in {d3e2} {
\foreach \a in {1,...,5} {
 \foreach \ea in {1,...,10} {
  \foreach \b in {1,...,5} {
  \foreach \eb in {1,...,10} {
  \newtoggle{closer}
   \ifnodedefined{d\a\ea}{
   	  \ifnodedefined{d\b\eb}{
   	  	\ifthenelse{\NOT \equal{\a}{\b} \OR \NOT \equal{\ea}{\eb} }{ 
		  \draw[->]
     		let \p1=(d\a\ea), \p2=(d\b\eb),
       		  \n3={\y2-\y1},
     		  \n2={\x2-\x1},
     		  \n1={veclen(\x2-\x1,\y2-\y1)} in
     		{\ifdim \n1 < 3.2cm
     		 \ifdim \n2 > 0.8cm
     		 (d\a\ea) -- (d\b\eb)
     		 \fi
     		 \fi};
     	}{}
   	  }{}
   }{}
  }
}}}}

\foreach \a in {1,...,5} {
 \foreach \ea in {1,...,10} {
 \pgfmathtruncatemacro\next{\ea+1}
 \ifnodedefined{d\a\next}{
 	\draw[->] (d\a\ea) -- (d\a\next);
}}}

\begin{scope}[on background layer]
    \filldraw[\colorFuture, line join=round, line width=1cm] plot coordinates{(d32) (d44) (d45) (d53) (d46) (d34) (d25) (d25) (d33) (d32)}--cycle;
    \filldraw[\colorPast, line join=round, line width=1cm] plot coordinates{(d32) (d42.south) (d51) (d41) (d21) (d11) (d22) (d32)}--cycle;
    \filldraw[\colorPresent, line join=round, line width=1cm] plot coordinates{(d32.north west)(d32.north east) (d32.south east) (d32.south west)}--cycle;

\end{scope}

\draw[<->] (0,0) -- node[xshift=-1cm]{Devices} (0,7);
\draw[->](0,0) -- node[yshift=-0.6cm]{Time} (10,0);
\end{tikzpicture}
}
\caption{Example of an event structure modelling a distributed system execution. Nodes labelled by $\eventId_k^\deviceId$ denote the $k$-th round of device $\deviceId$. The yellow background highlights a reference event, from which its past (green) and future (blue) are identified through the causal relationship implied by the arrows denoting neighbour events.}
\label{fig:event-structure-example}
\end{figure}

The semantic\revB{s} of a system execution can be expressed as an event structure (see Figure \ref{fig:event-structure-example}),
 where events $\eventId$ denote whole sense--compute--act rounds,
 and arrows between events denote that certain source events have provided inputs (i.e., messages) to target events.
 In particular, if event $\eventId'$ is connected with an arrow to $\eventId$, we say that $\eventId'$ is a \emph{neighbour} of $\eventId$, denoted $\eventId' \neigh \eventId$.
 We denote with $\neighset(\eventId)$ the set of all \emph{neighbours} of $\eventId$, with $\devof(\eventId)$ the device where event $\eventId$ happens (i.e., where it is executed)\extLmcs{,
 and with $\eprec(\eventId)$ the empty or singleton set with only the previous event of $\eventId$ at the same device.
}

Programming such systems 
 ultimately means defining the control rules
 that specify how the local context at each event
 is mapped to the messages to be sent to neighbour events.

\subsection{XC key data type: Neighbouring Values} \label{ssec:data}

In XC, we distinguish two types of values.
The \emph{Local} values $\lvalue$ include classic atomic and structured types $\ltype$ such as int, float, string, and list.
The neighbouring values (\emph{\excvalues}) are instead
maps $\underline{\anyvaluealt}$ from device identifiers $\deviceId_i$ to corresponding local values $\lvalue_i$, with an additional local value $\lvalue$ that acts as a \emph{default}:
\[\underline{\anyvaluealt} = \lvalue[\deviceId_1 \mapsto \lvalue_1, \ldots, \deviceId_n \mapsto \lvalue_n]\]
A \excvalue\ specifies a (set of) values received from or sent to neighbours: \revA{received values are gathered into \excvalues{}, then can be locally processed, and the final resulting \excvalue{} can be interpreted as messages to be sent back to neighbours.}
The devices with an associated entry in the \excvalue{} are thus
typically a (small) subset of all devices, namely those that are close-enough to the current device, and which are \revB{assumed to
work} correctly.

The default is used when a value is not available for some neighbour $\deviceId'$, e.g., because $\deviceId'$ has just been switched on and has not yet produced a value, or because it has just moved close enough to the current device $\deviceId$ to become one of its neighbours.
The notation above should therefore read as ``the {\excvalue} $\underline{\anyvaluealt}$ is $\lvalue$
everywhere (i.e., for all neighbours) except for devices $\deviceId_1, \ldots, \deviceId_n$
which have values $\lvalue_1, \ldots, \lvalue_n$''.

To exemplify \excvalues, in Figure~\ref{fig:event-structure-example},
upon waking up for computation $\eventId_2^3$, device $\deviceId_3$ may process a \excvalue\
$\underline{\anyvaluealt} = 0[\deviceId_4\!\mapsto\!1,\deviceId_3\!\mapsto\!2,\deviceId_2\!\mapsto\!3]$,
corresponding to the messages carrying scalar values 1, 2, and 3  received when asleep
from $\deviceId_4$, $\deviceId_3$ (i.e., \emph{itself} at the previous round), and $\deviceId_2$. The entries for all other (neighbour) devices default
to $0$.
After the computation,  $\deviceId_2$ may send out the messages denoted by the \excvalue\
$\underline{\anyvaluealt}' =  0[\deviceId_4\!\mapsto\!5,\deviceId_3\!\mapsto\!6]$; so that $5$ is sent to $\deviceId_4$, $6$ is sent to $\deviceId_3$, and $0$ is sent to every other device, such as a newly-connected device.
For convenience, we may use the notation $\underline{\anyvaluealt}(\deviceId')$ for the local value (specific or default) associated with $\deviceId'$ by $\underline{\anyvaluealt}$.

Note that a local value $\lvalue$ can be naturally converted to {a \excvalue} $\lvalue[]$ where it is the default value for every device.
Except for clarity, thus local values and \excvalues\ can be treated uniformly.
Functions on local values are implicitly lifted to \excvalues, by applying them on the maps' content pointwise.
For example, if $\underline{\anyvaluealt}_1$ assigns value $2$ to $\deviceId_3$ and $\underline{\anyvaluealt}_2$ assigns default value $1$ to $\deviceId_3$, then $\underline{\anyvaluealt}_3 = \underline{\anyvaluealt}_1 \cdot \underline{\anyvaluealt}_2$ shall assign value $2 \cdot 1 = 2$ to $\deviceId_3$.
\extLmcs{In \sysname{}, we also lift to \excvalues\ the \emph{multiplexer} operator $\muxK(\lvalue_1, \lvalue_2, \lvalue_3)$, which returns $\lvalue_2$ if $\lvalue_1$ is $\truevalue$ and $\lvalue_3$ otherwise.}

A fundamental operation on \excvalues\ is provided by the built-in function $\foldK(f: (A,B) \to A, \underline{\anyvaluealt}: \tfieldOf{B} , \lvalue: A): A$. As suggested by the name, the function folds over a \excvalue\ $\underline{\anyvaluealt}$, i.e., starting from a base local value $\lvalue$ it repeatedly applies function $f$ to neighbours' values in $\underline{\anyvaluealt}$, excluding the value for the current device.
For instance, if $\deviceId_2$ with set of neighbours $\bp{\deviceId_1, \deviceId_3, \deviceId_4}$ performs a $\foldK$ operation $\foldK(*, \underline{\anyvaluealt}, 1)$, the output will be $1 \cdot \underline{\anyvaluealt}(\deviceId_1) \cdot \underline{\anyvaluealt}(\deviceId_3) \cdot \underline{\anyvaluealt}(\deviceId_4)$.
Note that, as \excvalues\ are {\em unordered maps}, it is sensible to assume that $f$ is associative and commutative.

Two built-ins on \excvalues\ act on the value associated with the current (self) device:
\begin{itemize}
\item $\selfK(\underline{\anyvaluealt} : \tfieldOf{\ltype}) : \ltype \;\;$ returns the local value $\underline{\anyvaluealt}(\deviceId)$ in $\underline{\anyvaluealt}$ for the self device $\deviceId$
\item $\updateSelfK(\underline{\anyvaluealt} : \tfieldOf{\ltype}, \lvalue : \ltype) : \tfieldOf{\ltype} \;\;$ returns an \excvalue\ where the value for $\deviceId$ is set to $\lvalue$.
\end{itemize}

\extLmcs{
Furthermore, \sysname features a single communication primitive $\pExchange{\e_i}{\underline{\nname}}{\e_r}{\e_s}$ which is evaluated as follows.
\begin{itemize}
\item the device computes the local value $\lvalue_i$ of $\e_i$ (the \emph{initial} value);
\item to evaluate the function provided as second argument, it substitutes variable $\underline{\nname}$ with the \excvalue\ $\underline{\anyvaluealt}$ of messages
received from the neighbours for this exchange, using $\lvalue_i$ as default;
\item the exchange expression as a whole returns the (neighbouring or local) value $\anyvalue_r$ from the evaluation of $\e_r$;
\item the second expression of the argument function's output, $\e_s$, evaluates to a \excvalue\ $\underline{\anyvaluealt}_s$
consisting of local values to be sent to neighbour devices $\deviceId'$,
that will use their corresponding $\underline{\anyvaluealt}_s(\deviceId')$ as soon as
they wake up and perform their next execution round.
\end{itemize}
Since expressions $\anyvalue_r$  and $\anyvalue_s$  often coincide, we also provide $\sExchange{\e_i}{\underline{\nname}}{\e}$ as a shorthand for $\pExchange{\e_i}{\underline{\nname}}{\e}{\e}$.
This construct abstracts the general concept of message exchange, and is sufficiently expressive to allow other communication patterns to be expressed by it.
If a program executes multiple exchange-expressions, \sysname ensures through \emph{alignment} that the messages are dispatched to corresponding
exchange-expressions across rounds. The detailed workings of this process will be given in \Cref{ssec:opsem}.
}

\begin{figure}
\newcommand{\toAbove}{1.5cm}
\newcommand{\toRight}{1cm}
\centering
\adjustbox{width=0.8\columnwidth}{
\begin{tikzpicture}[evlabel/.style={color=red!70!black,font=\bfseries}]
\node[] (origin) at (0,0) {};

\foreach \dev/\evs/\offx/\freq in {%
1/3/ -0.8 / 3.5,%
2/5/ -0.2 / 2.3,%
3/4/ 0.3 / 2.6,%
4/6/ -0.2 / 1.8,%
5/3/ -0.8 / 3.8}
  \foreach \ev in {1,...,\evs}
        \node[present,font=\Large] (d\dev\ev) [above right = \dev*\toAbove and \ev*\toRight*\freq,xshift=-1cm+\offx*2cm,yshift=-1cm] {$\eventId_\ev^\dev$}; 

\newcommand{\nlabel}[2]{
  \foreach \lbl [count=\lbli] in {#2}
    \node [evlabel] (ld#1\lbli) [above=0cm of d#1\lbli] {$\lbl$};
}
\nlabel{5}{\infty, 1, 1}
\nlabel{4}{\infty, 0, 0,0,0,0}
\nlabel{3}{\infty, 1, 1, 1}
\nlabel{2}{\infty, 1, 3 ,\infty,2}
\nlabel{1}{\infty, 2, 4}

\makeatletter
\long\def\ifnodedefined#1#2#3{%
    \@ifundefined{pgf@sh@ns@#1}{#3}{#2}%
}
\makeatother
\foreach \ref in {d3e2} {
\foreach \a in {1,...,5} {
 \foreach \ea in {1,...,10} {
  \foreach \b in {1,...,5} {
  \foreach \eb in {1,...,10} {
  \newtoggle{closer}
   \ifnodedefined{d\a\ea}{
   	  \ifnodedefined{d\b\eb}{
   	  	\ifthenelse{\NOT \equal{\a}{\b} \OR \NOT \equal{\ea}{\eb} }{ 
		  \draw[->]
     		let \p1=(d\a\ea), \p2=(d\b\eb),
       		  \n3={\y2-\y1},
     		  \n2={\x2-\x1},
     		  \n1={veclen(\x2-\x1,\y2-\y1)} in
     		{\ifdim \n1 < 3.2cm
     		 \ifdim \n2 > 0.8cm
     		 (d\a\ea) -- (d\b\eb)
     		 \fi
     		 \fi};
     	}{}
   	  }{}
   }{}
  }
}}}}

\foreach \a in {1,...,5} {
 \foreach \ea in {1,...,10} {
 \pgfmathtruncatemacro\next{\ea+1}
 \ifnodedefined{d\a\next}{
 	\draw[->] (d\a\ea) -- (d\a\next);
}}}

\draw[<->] (0,0) -- node[xshift=-1cm]{Devices} (0,7);
\draw[->](0,0) -- node[yshift=-0.6cm]{Time} (10,0);
\end{tikzpicture}
}
\caption{\revB{
Example of a gradient computation (cf. \autoref{ex:gradient}) on the same event structure of \Cref{fig:event-structure-example}.
Assume that device 4 becomes source at its 2nd round, i.e., at event $\eventId_2^4$,
 and that neighbouring events are at a unit distance, i.e.,
 $D_\eventId(d(\eventId)')=1$ for all $\eventId \neigh \eventId'$.
Then, the output of the gradient computation at some event
 is shown with a red label above that event.
In each event,
  the gradient value is computed by
  taking the gradient values at past neighbour events (excluding past events of the same device),
  adding the corresponding distance
  (which we assume any device is able to estimate locally),
  and selecting the minimum of them.
For instance, $\eventId_5^2$ gathers $[\eventId_3^3 \mapsto 1, \eventId_3^1 \mapsto 4]$, adds neighbouring distances to get $[\eventId_3^3 \mapsto 2, \eventId_3^1 \mapsto 5]$, and selects the minimum value, i.e., $2$.
If there are no neighbouring events (as in $\eventId_1^4$ or $\eventId_4^2$), then $\infty$ is produced as the result of minimisation.
}}
\label{fig:event-structure-gradient}
\end{figure}

\begin{exa}[Gradient]\label{ex:gradient}
  A simple computation that can be executed on event structures is
  a distributed gradient algorithm, which computes everywhere in a network of devices
  the minimum distance from each device to its closest \emph{source} device.

  The \emph{Adaptive Bellman-Ford (ABF)} algorithm~\cite{DBLP:journals/tac/MoDB19} computes the gradient by assuming that each device is able to get (an estimation of) its distance to each one of its neighbours.
Let $S$ denote the set of sources, \revB{$D_\eventId(\deviceId')$ be the distance between devices $\deviceId = \devof(\eventId)$ and $\deviceId'$ as estimated in event $\eventId$}, and $g(\eventId)$ be the resulting gradient value computed at $\eventId$. Then, the ABF algorithm on event structures can be expressed as follows:
$$
g(\eventId) = \begin{dcases}
\min_{\eventId' \in \neighset(\eventId)\setminus\eprec(\eventId)} \{ g(\eventId') +  D_{\eventId}(\devof(\eventId')) \} & \devof(\eventId) \notin S
\\
0 & \devof(\eventId) \in S
\end{dcases}
$$
\revB{The gradient value is set to zero on source devices. For other devices, it is set by considering each neighbour $\eventId'$ of $\eventId$ excluding the current device itself: $\neighset(\eventId)\setminus\eprec(\eventId)$. For each such neighbour, the algorithm selects the minimum distance as computed through it, by summing up the estimates of the distance $g(\eventId')$ from the closest source to the neighbour with the distance $D_{\eventId}(\devof(\eventId'))$ from the neighbour to the current event.}
By running such function in each event, the nodes will tend to adjust their gradient value towards the ``correct'' distance value. \revB{Figure \ref{fig:event-structure-gradient} presents a sample execution of this function on an event structure.}

\revB{In \sysname, the gradient function just described can be written as follows:}
\begin{minipage}{\textwidth}
\begin{lstlisting}[language=xc,mathescape]
def gradient(isSource) {
  exchange( $\infty$, (n) => nfold(min, n + nbrDist(), mux(isSource, 0, $\infty$)) )
}
\end{lstlisting}
\end{minipage}
\revB{In the code above, we assume that the built-in function \texttt{nbrDist} returns in each event $\eventId$ the \excvalue\ $\evalue{0}{\overline\deviceId}{D_\eventId(\overline\deviceId)}$, where $\overline\deviceId = \neighset(\eventId)\setminus\eprec(\eventId)$. Furthermore, the built-in function $\foldK(\minK, \underline\anyvaluealt, \lvalue)$ returns the minimum across $\lvalue$ and each $\underline\anyvaluealt(\deviceId')$ for a neighbour $\deviceId' \neq \deviceId$ excluding the current device $\deviceId = \devof(\eventId)$.}
\end{exa}

\begin{figure}
\centerline{\framebox[\linewidth]{$\begin{array}{l}
	\textbf{Syntax:} \\
	\begin{array}{c@{\hspace{3pt}}c@{\hspace{3pt}}l@{\hspace{18mm}}r}
		\e & \BNFcce &
									\xname
				\, \BNFmid \, \funK \xname(\overline\xname) \{ \e \}
				\, \BNFmid \, \e(\overline\e)
				\, \BNFmid \, \valK \; \xname = \e; \e
				\, \BNFmid \, \lvalue
				\, \BNFmid \, \anyvaluealt
		& {\footnotesize \mbox{expression}}
		\\[3pt]
		\anyvaluealt & \BNFcce & \evalue{\lvalue}{\overline\deviceId}{\overline\lvalue}
		& {\footnotesize \mbox{\excvalue}}
		\\[3pt]
		\lvalue & \BNFcce & \bname \, \BNFmid \, \funK \xname(\overline\xname) \{ \e \} \, \BNFmid \, \cname(\overline\lvalue)
		& {\footnotesize \mbox{local literal}}
		\\[3pt]
		\bname & \BNFcce & \exchangeK \, \BNFmid \, \foldK \, \BNFmid \, \selfK \, \BNFmid \, \updateSelfK \, \BNFmid \, \uidK \, \BNFmid \, \ldots
		& {\footnotesize \mbox{built-in function}}
	\end{array} \\
		\hline\\[-10pt]
	\textbf{Free variables of an expression:} 
	\\
	\begin{array}{ll}
		\FV{\xname} = \{ \xname \}  \quad \FV{\lvalue} = \FV{\anyvaluealt} = \emptyset
		&
		\FV{\funK \xname_0(\xname_1,\ldots,\xname_n) \{ \e \}} = \FV{\e} \setminus \bp{\xname_0,\ldots,\xname_n}
		\\[2pt]
		\medmuskip=1mu
		\thinmuskip=1mu
		\thickmuskip=1mu
		\nulldelimiterspace=0pt
		\scriptspace=0pt
		\FV{\e_0(\e_1,\ldots,\e_n)} \,=\, \bigcup_{i=0 \ldots n} \FV{\e_i}
		&
		\FV{\valK \; \xname = \e; \e'} = \FV{\e} \cup \FV{\e'} \setminus \bp{\xname}
	\end{array}\\
	\hline\\[-10pt]
	\textbf{Syntactic sugar:} \\
	\begin{array}{c@{\hspace{3pt}}c@{\hspace{3pt}}l}
		(\overline\xname) \toSymK \e & \BNFcce & \funK \yname(\overline\xname) \{ \e \} \textit{ where } \yname \textit{ is a fresh variable}
		\\[3pt]
		\defK \xname(\overline\xname) \{ \e \} & \BNFcce & \valK \xname = \funK \xname(\overline\xname) \{ \e \};
		\\[3pt]
		\ifK (\e) \{\e_\top\} \elseK \{\e_\bot\} & \BNFcce & \muxK(\e, () \toSymK \e_\top, () \toSymK \e_\bot)()
	\end{array}
\end{array}$}}
\caption{\extLmcs{Syntax (top), free variables (middle) and   syntactic sugar (bottom) for \sysname\ expressions.}} \label{fig:pfc:syntax}
\end{figure}

\extLmcsStart
\subsection{Syntax} \label{ssec:pfc:syntax}

\Cref{fig:pfc:syntax} (top) shows the formal syntax of \sysname.
As in~\cite{FJ}, the overbar notation indicates a (possibly empty) sequence of elements, e.g., $\overline{x}$ is short for $x_1, \ldots, x_n$ $(n\ge 0)$.
The syntax is that of a standard functional language, with no peculiar features for distribution, that are instead apparent in the semantics.
An \sysname \emph{expression} $\e$ can be either:
\begin{itemize}
	\item
	a \emph{variable} $\xname$; 
	\item
	a (possibly recursive) \emph{function} $\funK \xname(\overline\xname) \{ \e \}$, which may have free variables;
	\item
	a \emph{function call} $\e(\overline\e)$;
	\item
	a \emph{let-style} expression $\valK \xname = \e; \e$;
	\item
	a \emph{local literal} $\lvalue$, that is either a built-in function $\bname$, a defined function $\funK \xname(\overline\xname) \{ \e \}$ \emph{without} free variables, or a data constructor $\cname$ applied to local literals (possibly none);
	\item
	an \emph{\excvalue}\ $\anyvaluealt$, as described in \Cref{ssec:data}.
\end{itemize}
\sysname can be typed using standard higher-order let-polymorphism, \revB{as shown in the following section}, without distinguishing between types for local and neighbouring values. This will reflect in the semantics by having constructs and built-in functions of the language accepting \excvalues\ for their arguments, and by having implicit promotion of local values $\lvalue$ to \excvalues\ $\lvalue[]$. Free variables are defined in a standard way (\Cref{fig:pfc:syntax}, middle), and an expression $\e$ is \emph{closed} if $\FV{\e}=\emptyset$. Programs are closed expressions without \excvalues\ as sub-expressions: \excvalues\ only arise in computations, and are the only values produced by evaluating programs.
The minimal syntax presented is often extended through syntactic sugar to include infix operators, omitted parenthesis in 0-ary constructors, and other non-trivial encoding described in \Cref{fig:pfc:syntax} (bottom).


\subsection{\revB{Typing}} \label{ssec:typing}

\revB{
\Cref{fig:pfc:typing} shows a classic Hindley-Milner type system~\cite{DBLP:conf/popl/DamasM82} for \sysname.
A type $\type$ can be:
\begin{itemize}
	\item
	a \emph{type variable} $\tvar$;
	\item
	a (recursive) \emph{data type} $\ltypeOf{\overline\type}$, consisting of a \emph{parametric type name}  $\ltypefun$ of arity $n\ge 0$, applied to $n$ types $\type_1,...,\type_n$ (possibly zero);
	\item
	or a \emph{function type} $(\overline\type) \rightarrow \type$.
\end{itemize}
Given that local values and \excvalues\ are treated uniformly in the semantics of \sysname, and only \excvalues\ result from evaluating closed expressions, no special types are needed for \excvalues: each \excvalue\ takes on the (classic) type of its messages. The underlining notation for \excvalue\ types and variables used before is simply for readability and not part of the type system itself.
The set of type variables in $\type$ is denoted by $\FTV{\type}$. Polymorphism in functions and data constructors is supported by type schemes $\typescheme$ of the form $\forall \overline{\tvar}.\type$, where $\overline{\tvar}$ are free in $\type$, representing all types obtained by substituting $\overline{\tvar}$ with types $\overline{\type}$, as described by the type scheme instantiation relation $\tsinsrel$.
}

\begin{figure}[tb]
\revB{\centerline{\framebox[\linewidth]{$\begin{array}{l}
	\textbf{Types, type schemes, typing environments  and instantiation:} \\
	\begin{array}{c@{\hspace{3pt}}c@{\hspace{3pt}}l@{\hspace{-4mm}}r @{\hspace{10mm}} c@{\hspace{3pt}}c@{\hspace{3pt}}l@{\hspace{4mm}}r}
		\type & \BNFcce & \tvar \, \BNFmid \, \ltypeOf{\overline\type} \, \BNFmid \, (\overline\type) \rightarrow \type
		& {\footnotesize \mbox{type}}
		&
		\aname & \BNFcce & \xname \, \BNFmid \, \bname \, \BNFmid \,  \cname
		& {\footnotesize \mbox{assumption subject}}
		\\[4pt]
		\typescheme & \BNFcce & \forall \overline{\tvar}.\type
		& {\footnotesize \mbox{type scheme}}
		&
		\TtypEnv & \BNFcce & \emptyseq \, \BNFmid \, \TtypEnv, \aname : \typescheme
		& {\footnotesize \mbox{typing environment}}
		\\[4pt]
		\forall\overline{\tvar}.\type 	 & \tsinsrel  & \type\,[\overline{\tvar}:=\overline{\type}]   & {\footnotesize
		\mbox{instantiation}}
	\end{array} \\
	\hline\\[-10pt]
	\textbf{Expression typing:} \hspace{85mm} \boxed{\expTypJud{\TtypEnv}{\e}{\type}} \\
	\begin{array}{c}
		\surfaceTyping{T-ASS}{ ~
			\TtypEnv(\aname) \tsinsrel \type \text{ for } \aname = \xname \text{ or } \bname
		}{
			\expTypJud{\TtypEnv}{\aname}{\type}
		}
		\qquad
			\surfaceTyping{T-LIT}{ ~
			\TtypEnv(\cname) \tsinsrel (\overline\type)\rightarrow\type \quad
			\expTypJud{\TtypEnv}{\overline\lvalue}{\overline{\type}}
		}{
			\expTypJud{\TtypEnv}{\cname(\overline\lvalue)}{\type}
		}
		\skiptransition
		\surfaceTyping{T-NVAL}{ ~
			\expTypJud{\TtypEnv}{\lvalue,\overline\lvalue}{\type} \qquad
			\overline\deviceId \text{ distinct}
		}{
			\expTypJud{\TtypEnv}{\evalue{\lvalue}{\overline\deviceId}{\overline\lvalue}}{\type}
		}
		\qquad\qquad
		\surfaceTyping{T-APP}{ ~
			\expTypJud{\TtypEnv}{\e}{(\overline\type)\rightarrow\type} \qquad
			\expTypJud{\TtypEnv}{\overline\e}{\overline{\type}}
		}{
			\expTypJud{\TtypEnv}{\e(\overline\e)}{\type}
		}
		\skiptransition
		\surfaceTyping{T-FUN}{ ~
			\expTypJud{\TtypEnv, \xname : \forall \emptyseq. (\overline\type) \to \type, \overline\xname : \forall \emptyseq. \overline\type}{\e}{\type} \qquad
		}{
			\expTypJud{\TtypEnv}{\funK \xname(\overline\xname)\{\e\}}{(\overline\type) \to \type}
		}
		\skiptransition
		\surfaceTyping{T-VAL}{ ~
			\expTypJud{\TtypEnv}{\e_1}{\type_1} \qquad
			\overline\tvar = \FTV{\type_1} \qquad
			\expTypJud{\TtypEnv,\xname : \forall \overline{\tvar}.\type_1}{\e_2}{\type_2} \qquad
		}{
			\expTypJud{\TtypEnv}{\valK \xname = \e_1; \e_2}{\type_2}
		}
	\end{array} \\
\end{array}$}}}
\caption{\revB{Typing of \sysname\ expressions.}} \label{fig:pfc:typing}
\end{figure}

\revB{
A \emph{typing environment} $\TtypEnv$ is a set of \emph{assumptions} $\aname : \typescheme$ where the $\aname$ can be a variable, built-in function or constructor.
Sensor types $\sname$ are assumed to be of the form $() \to \type$ with $\FTV{\type} = \emptyset$. For program typing, an \emph{initial typing environment} $\TtypEnv_0$ defines a type scheme for each available constructor and built-in function. This environment is extended with further assumptions for bounded variables as they appear within sub-expressions of the program. We denote by $\TtypEnv(\aname)$ the unique type scheme assigned to $\aname$ in $\TtypEnv$.

Expression typing is specified by judgement $\expTypJud{\TtypEnv}{\e}{\type}\,$
meaning ``expression $\e$ has type $\type$ under assumptions $\TtypEnv$''.
Following~\cite{FJ}, we expand multiple overbars together, so for instance $\expTypJud{\TtypEnv}{\overline\e}{\overline\type}$ represents $\expTypJud{\TtypEnv}{\e_1}{\type_1}, \cdots, \expTypJud{\TtypEnv}{\e_n}{\type_n}$.
Typing rules are all syntax-directed and mostly standard, except for \ruleNameSize{[T-NVAL]} that is \sysname{}-specific. This rule ensures that all messages $\lvalue, \overline\lvalue$ within an \excvalue\ share the same type $\type$ and that $\overline\deviceId$ has no duplicate entries.
}


\subsection{Semantics} \label{ssec:opsem}

The \sysname\ semantics is defined as
\emph{(i)} a big-step operational \emph{device semantics}, providing a formal account of the computation of a device within one round; and
\emph{(ii)} a denotational \emph{network semantics}, formalising how different device rounds communicate.

\begin{figure}[t]
\centerline{\framebox[\linewidth]{$\begin{array}{l}
	\textbf{Auxiliary definitions:} \\ \!\!\!
	\begin{array}{ccl@{\hspace{6mm}}r @{\hspace{35mm}} l@{\hspace{6mm}}r}
		\vtree & \BNFcce & {\mkvt{}{\overline{\vtree}} \; \BNFmid} \; \mkvt{\anyvaluealt}{ {\overline\vtree} } &   {\footnotesize \mbox{value-tree}} &
		\senstate & {\footnotesize \mbox{sensor state}}
		\\
		\Trees & \BNFcce & \envmap{\overline{\deviceId}}{\overline{\vtree}} &   {\footnotesize \mbox{value-tree environment}} &
		\deviceId & {\footnotesize \mbox{device identifier}} \\[5pt]
	\end{array} \\ \!\!\!
	\begin{array}{l}
		{\piIof{i}{\mkvt{}{\vtree_1,\ldots,\vtree_n}} = \vtree_i} \qquad
		{\piIof{i}{\mkvt{\anyvaluealt}{\vtree_1,\ldots,\vtree_n}} = \vtree_i} \qquad
		{\piIof{i}{\envmap{\overline\deviceId}{\overline\vtree}} = \envmap{\overline\deviceId}{\piIof{i}{\overline\vtree}}}
		\\[2pt]
		{\proj{\envmap{\overline\deviceId}{\overline\vtree}}{\funvalue} = \bp{\envmap{\deviceId_i}{\vtree_i} ~\mid~ \vtree_i = \mkvt{\anyvaluealt}{\overline\vtree'}, \; \nameOf(\anyvaluealt(\deviceId_i)) = \nameOf(\funvalue)}}
		\begin{array}{l}
			\nameOf(\bname) = \bname \\
			\nameOf(\funKname \xname(\overline{\xname}) \{ \e \}) = \name
		\end{array} \\[10pt]
	\end{array} \\
	\hline\\[-10pt]
	\textbf{Evaluation rules:} \hspace{77mm} \boxed{\eqhlbsopsem{\deviceId}{\Trees}{\senstate}{\e}{\anyvaluealt}{\vtree}}
	\\ \qquad
	\begin{array}{c}
		\begin{array}{c}
			\nullsurfaceTyping{E-NBR}{
				\!\!\! \eqhlbsopsem{\deviceId}{\Trees}{\senstate}{\anyvaluealt}{\anyvaluealt}{\mkvt{}{}} \!\!\!\!\!
			}
			\\[-3pt]
			\nullsurfaceTyping{E-LIT}{
				\!\!\! \eqhlbsopsem{\deviceId}{\Trees}{\senstate}{\lvalue}{\lvalue[]}{\mkvt{}{}} \!\!\!\!\!
			}
		\end{array}
		\qquad
		\surfaceTyping{E-VAL}{
			\begin{array}{l}
				\eqhlbsopsem{\deviceId}{\piI{1}(\Trees)}{\senstate}{\e_1}{\anyvaluealt_1}{\vtree_1} \\
				\eqhlbsopsem{\deviceId}{\piI{2}(\Trees)}{\senstate}{\applySubstitution{\e_2}{\substitution{\xname}{\anyvaluealt_1}}}{\anyvaluealt_2}{\vtree_2}
			\end{array}
		}{
			\eqhlbsopsem{\deviceId}{\Trees}{\senstate}{\valK \xname = \e_1; \e_2}{\anyvaluealt_2}{\mkvt{}{\vtree_1, \vtree_2}} \!\!\!\!\!
		}
		\skiptransition
		\surfaceTyping{E-APP}{
		\begin{array}{l}
			\eqhlbsopsem{\deviceId}{\piI{i+1}(\Trees)}{\senstate}{\e_i}{\anyvaluealt_i}{\vtree_i} \quad
			\text{for all}\; i \in 0, \ldots, n
			\\
			\eqhlauxopsem{\deviceId}{\piI{n+2}(\proj{\Trees}{\funvalue})}{\senstate}{\funvalue(\anyvaluealt_1, \ldots, \anyvaluealt_n)}{\anyvaluealt_{n+1}}{\vtree_{n+1}} \;
			\text{where } \funvalue = \anyvaluealt_0(\deviceId)
		\end{array} \!\!\!\!
		}{
			\eqhlbsopsem{\deviceId}{\Trees}{\senstate}{\e_0(\e_1,\ldots,\e_n)}{\anyvaluealt_{n+1}}{\mkvt{\funvalue[]}{\vtree_0,\ldots,\vtree_{n+1}}}
		}
		\\[20pt]
	\end{array} \\
	\hline\\[-10pt]
	\textbf{Auxiliary evaluation rules:} \hspace{54mm} \boxed{\eqhlauxopsem{\deviceId}{\Trees}{\senstate}{\funvalue(\overline\anyvaluealt)}{\anyvaluealt}{\vtree}} \\
	\begin{array}{c}
		\surfaceTyping{A-FUN}{
		\begin{array}{l}
			\eqhlbsopsem{\deviceId}{\Trees}{\senstate}{\applySubstitution{\e}{\substitution{\xname}{\funKname \xname(\overline\xname)\{\e\}}, \substitution{\overline\xname}{\overline\anyvaluealt}}}{\anyvaluealt}{\vtree}
		\end{array} \!\!\!\!
		}{
			\eqhlauxopsem{\deviceId}{\Trees}{\senstate}{\funKname \xname(\overline\xname)\{\e\}(\overline\anyvaluealt)}{\anyvaluealt}{\vtree}
		}
		\qquad
		\nullsurfaceTyping{A-UID}{
			\eqhlauxopsem{\deviceId}{\Trees}{\senstate}{\uidK()}{\deviceId}{\mkvt{}{}}
		}
		\skiptransition
		\surfaceTyping{A-XC}{
			\begin{array}{l}
				\Trees = \envmap{\overline\deviceId}{\mkvt{\overline\anyvaluealt}{\ldots}} \quad
				\anyvaluealt = \anyvaluealt_i[\envmap{\overline\deviceId}{\overline\anyvaluealt(\deviceId)}] \quad
				\funvalue = \anyvaluealt_f(\deviceId)
				\\
				\eqhlauxopsem{\deviceId}{\piI{1}(\Trees)}{\senstate}{\funvalue(\anyvaluealt)}{(\anyvaluealt_r, \anyvaluealt_s)}{\vtree}
			\end{array}
			\!\!\!\!\!
		}{
			\eqhlauxopsem{\deviceId}{\Trees}{\senstate}{\exchangeK(\anyvaluealt_i, \anyvaluealt_f)}{\anyvaluealt_r}{\mkvt{\anyvaluealt_s}{\vtree}}
		}
		\qquad
		\nullsurfaceTyping{A-SELF}{
			\eqhlauxopsem{\deviceId}{\Trees}{\senstate}{\selfK(\anyvaluealt)}{\anyvaluealt(\deviceId)}{\mkvt{}{}}
		}
		\skiptransition
		\surfaceTyping{A-FOLD}{
			\begin{array}{l}
				\Trees = \envmap{\deviceId_1}{\vtree_1}, \ldots, \envmap{\deviceId_n}{\vtree_n}
				\qquad
				\lvalue_0 = \anyvaluealt_3(\deviceId)
				\qquad
				\funvalue = \anyvaluealt_1(\deviceId)
				\\
				\eqhlauxopsem{\deviceId}{\emptyset}{\senstate}{\funvalue(\lvalue_{i-1}[], \anyvaluealt_2(\deviceId_i)[])}{\lvalue_i[]}{\vtree}
				\text{ if } \deviceId_i \neq \deviceId \text{ else } \lvalue_i = \lvalue_{i-1}
			\end{array}
			\!\!\!\!\!
		}{
			\eqhlauxopsem{\deviceId}{\Trees}{\senstate}{\foldK(\anyvaluealt_1, \anyvaluealt_2, \anyvaluealt_3)}{\lvalue_n[]}{\mkvt{}{}}
		}
		\qquad
		\cdots
	\end{array} \!\!\!\!
\end{array}$}}
\caption{\extLmcs{Device big-step operational semantics of \sysname}.} \label{fig:pfc:opsem-dev}
\end{figure}

\paragraph*{Device semantics.} \Cref{fig:pfc:opsem-dev} presents the device semantics, formalised by judgement $\bsopsem{\deviceId}{\Trees}{\senstate}{\e}{\anyvaluealt}{\vtree}$, to be read as ``expression $\e$ evaluates to \excvalue\ $\anyvaluealt$ and value-tree $\vtree$ on device $\deviceId$ with respect to sensor values $\senstate$ and value-tree environment $\Trees$'', where:
\begin{itemize}
\item
    $\anyvaluealt$ is called the \emph{result} of $\e$;
	\item
	$\vtree$ is an ordered tree with \excvalues\ on some nodes (cf.~\Cref{fig:pfc:opsem-dev} (top)), representing messages to be sent to neighbours by tracking the \excvalues\ produced by exchange-expressions in $\e$, and the stack frames of function calls;
	\item
	$\Trees$ collects the (non expired) value-trees received by the most recent firings of neighbours of $\deviceId$, as a map $\envmap{\deviceId_1}{\vtree_1}$, $\ldots$, $\envmap{\deviceId_n}{\vtree_n}$ $(n \ge 0)$ from device identifiers to value-trees.
\end{itemize}
\revB{
The semantics of function application is expressed through an auxiliary evaluation judgement of similar form, $\eqhlauxopsem{\deviceId}{\Trees}{\senstate}{\funvalue(\overline\anyvaluealt)}{\anyvaluealt}{\vtree}$, to be read as ``expression $\funvalue(\overline\anyvaluealt)$ evaluates to \excvalue\ $\anyvaluealt$ and value-tree $\vtree$ on device $\deviceId$ with respect to sensor values $\senstate$ and value-tree environment $\Trees$''.
}
We assume every  function expression $\funK \xname(\overline\xname) \{\e\}$ occurring in the program is annotated with a unique name $\name$ before the evaluation starts. Then, $\name$ will be the name for the annotated function expression $\funKname \xname(\overline\xname) \{\e\}$, and $\bname$ the name for a built-in function $\bname$.

The syntax of value-trees and value-tree environments is in \Cref{fig:pfc:opsem-dev} (top).
The rules for judgement $\bsopsem{\deviceId}{\Trees}{\senstate}{\e}{\anyvalue}{\vtree}$ (\Cref{fig:pfc:opsem-dev}, middle) are standard for functional languages, extended to evaluate a sub-expression $\e'$ of $\e$ w.r.t. the value-tree environment $\Trees'$ obtained from $\Trees$ by extracting the corresponding subtree (when present) in the value-trees in the range of $\Trees$. This \emph{alignment} process is modelled by the auxiliary ``projection'' functions $\piI{i}$ (for any positive natural number $i$) (\Cref{fig:pfc:opsem-dev}, top). When applied to a value-tree $\vtree$, $\piI{i}$ returns the $i$-th sub-tree $\vtree_i$ of $\vtree$. When applied to a value-tree environment $\Trees$, $\piI{i}$ acts pointwise on the value-trees in $\Trees$.
The alignment process ensures that the value-trees in the environment $\Trees$ always correspond to the evaluation of the same sub-expression currently being evaluated. To ensure this match holds (as said before, of the stack frame and position in the AST), in the evaluation of a function application $\funvalue(\overline\anyvaluealt)$, the environment $\Trees$ is reduced to the smaller set $\proj{\Trees}{\funvalue}$ of trees which corresponded to the evaluation of a function with the same \emph{name}.

Rule \ruleNameSize{[E-NBR]} evaluates an \excvalue\ $\anyvaluealt$ to $\anyvaluealt$ itself and the empty value-tree.
Rule \ruleNameSize{[E-LIT]} evaluates a local literal $\lvalue$  to the \excvalue\ $\lvalue[]$ and the empty value-tree.
Rule \ruleNameSize{[E-VAL]} evaluates a val-expression, by evaluating the first sub-expression with respect to the first sub-tree of the environment obtaining a result $\anyvaluealt_1$, and then the second sub-expression with respect to the second sub-tree of the environment, after substituting the variable $\xname$ with $\anyvaluealt_1$.

Rule \ruleNameSize{[E-APP]} is standard eager function application: the function expression $\e_0$ and each argument $\e_i$ are evaluated
w.r.t. $\piIof{i+1}{\Trees}$ producing result $\anyvalue_i$ and value-tree $\vtree_i$. Then, the function application itself is demanded to the auxiliary evaluation rules, w.r.t. the last sub-tree of the trees corresponding to the same function: $\piIof{n+2}{\proj{\Trees}{\funvalue}}$.
The auxiliary rule \ruleNameSize{[A-FUN]} handles the application of fun-expression, which evaluates the body after replacing the arguments $\overline\xname$ with their provided values $\overline\anyvaluealt$, and the function name $\xname$ with the fun-expression itself \revB{to allow for recursion.}
Rules \ruleNameSize{[A-UID]} and \ruleNameSize{[A-SELF]} trivially encode the behaviour of the $\uidK$ and $\selfK$ built-ins.
Rule \ruleNameSize{[A-XC]} evaluates an exchange-expression, realising the behaviour described at the end of Section~\ref{ssec:data}. Notation $\evalue{\anyvaluealt_1}{\overline\deviceId}{\overline\anyvaluealt(\deviceId)}$ is used to represent the \excvalue\ $\anyvaluealt_1$ after the update for each $i$ of the message for $\deviceId_i$ with the custom message $\anyvaluealt_i(\deviceId)$. The result is fed as argument to function $\anyvaluealt_f$: the first element of the resulting pair is the overall result, while the second is used to tag the root of the value-tree.
Rule \ruleNameSize{[A-FOLD]} encodes the $\foldK$ operators. First, the domain of $\Trees$ is inspected, giving a (sorted) list $\deviceId_1, \ldots, \deviceId_n$. An initial local value $\lvalue_0$ is set to the ``self-message'' of the third argument. Then, a sequence of $\lvalue_i$ is defined, each by applying function $\anyvaluealt_1$ to the previous element in the sequence and the value $\anyvaluealt_2(\deviceId_i)$ (skipping $\deviceId$ itself). The final result $\lvalue_n$ is the result of the application, with empty value-tree.
Auxiliary rules for the other available built-in functions are standard, do not depend on the environment, hence have been omitted.

\begin{exa}[\revB{Gradient Semantics}]\label{ex:gradient:sem}
\revB{Let us consider the semantics of the gradient function presented in Example \ref{ex:gradient}, as used in the following program:}
\begin{lstlisting}[language=xc,mathescape]
def gradient(isSource) {
  exchange( $\infty$, (n) => nfold(min, n + nbrDist(), mux(isSource, 0, $\infty$)) )
}
gradient(sourceSensor())
\end{lstlisting}
\revB{After removing all syntactic sugar (described in \Cref{fig:pfc:syntax}), the program converts to:}
\begin{minipage}{\textwidth}
\begin{lstlisting}[language=xc,mathescape]
val gradient = fun gradient(isSource) {
  exchange( $\infty$, fun f(n) { nfold(min, n + nbrDist(), mux(isSource, 0, $\infty$)) } )
};
gradient(sourceSensor())
\end{lstlisting}
\end{minipage}
\revB{Consider the execution of such program in a given device $\deviceId$, with respect to sensor values $\senstate$ and value-tree environment $\Trees$.
The overall program is a val-expression, hence rule \ruleNameSize{[E-VAL]} applies. This rule demands the evaluation of the first sub-expression first (w.r.t. $\pi_1(\Trees)$), which is \lstinline[language=xc]|fun gradient...|. This is a local literal defining a function, so rule \ruleNameSize{[E-LIT]} applies, and produces an \excvalue\ without exceptions that corresponds to the function (and empty value-tree). Then, this resulting value is substituted for the variable \lstinline[language=xc]|gradient| in the last part of the program, obtaining:}
\begin{lstlisting}[language=xc,mathescape]
fun gradient(isSource) {
  exchange( $\infty$, fun f(n) { nfold(min, n + nbrDist(), mux(isSource, 0, $\infty$)) } )
}[](sourceSensor())
\end{lstlisting}
\revB{
According to rule \ruleNameSize{[E-VAL]}, this code has to be evaluated (w.r.t $\pi_2(\Trees)$). This is a function application, so rule \ruleNameSize{[E-APP]} applies. This evaluates the function first (w.r.t. $\pi_1(\pi_2(\Trees))$), which is already an \excvalue, so it evaluates to itself (and no value-tree) through rule \ruleNameSize{[E-NBR]}. Then, it evaluates the argument \lstinline[language=xc]|sourceSensor()| (w.r.t. $\pi_2(\pi_2(\Trees))$), which is a further function application. Without describing this further evaluation in details for brevity, it will ultimately produce a Boolean value $b$ that is extracted from the sensor status $\senstate$ through the auxiliary evaluation rule defined for the built-in function \lstinline[language=xc]|sourceSensor|.

As the last step of the overall function application, the auxiliary predicate is used, applying rule \ruleNameSize{[A-FUN]} to expand the function body (w.r.t. $\pi_3(\pi_2(\Trees))$). The substitution results in the following code to be evaluated:}
\begin{lstlisting}[language=xc,mathescape]
exchange( $\infty$, fun f(n) { nfold(min, n + nbrDist(), mux($b$[], 0, $\infty$)) } )
\end{lstlisting}
\revB{This is another function application where all expressions are local literals, so it resorts to rule \ruleNameSize{[E-APP]}, where the only interesting part is evaluating the application itself (w.r.t. $\Trees' = \pi_4(\pi_3(\pi_2(\Trees)))$) with the auxiliary rule \ruleNameSize{[A-XC]}. If we consider $\deviceId, \senstate, \Trees$ to be as in event $\eventId^2_5$ described in \Cref{fig:event-structure-gradient}, we have that $b = \bot$, $\deviceId = \deviceId_2$ and
$$\Trees' = \deviceId_1 \mapsto \mkvt{4[]}{\ldots}, \deviceId_2 \mapsto \mkvt{\infty[]}{\ldots}, \deviceId_3 \mapsto \mkvt{1[]}{\ldots}.$$
According to rule \ruleNameSize{[A-XC]}, first we compute a \excvalue\ combining received messages (using the first argument as default), which in this case is $\anyvaluealt = \infty[\deviceId_1 \mapsto 4, \deviceId_2 \mapsto \infty, \deviceId_3 \mapsto 1]$. This value is fed to the second argument, resorting in the function application:}
\begin{lstlisting}[language=xc,mathescape]
fun f(n) { nfold(min, n + nbrDist(), mux($\bot$[], 0, $\infty$)) }($\infty[\deviceId_1 \mapsto 4, \deviceId_2 \mapsto \infty, \deviceId_3 \mapsto 1]$)
\end{lstlisting}
\revB{Which after a further application of \ruleNameSize{[A-FUN]} translates to the evaluation of:}
\begin{lstlisting}[language=xc,mathescape]
nfold(min, $\infty[\deviceId_1 \mapsto 4, \deviceId_2 \mapsto \infty, \deviceId_3 \mapsto 1]$ + nbrDist(), mux($\bot$[], 0, $\infty$))
\end{lstlisting}
\revB{The evaluation proceeds similarly. Following the example in \Cref{fig:event-structure-gradient}, \lstinline[language=xc]|nbrDist()| evaluates to $1[]$,
which added to $\anyvaluealt$ becomes $\infty[\deviceId_1 \mapsto 5, \deviceId_2 \mapsto \infty, \deviceId_3 \mapsto 2]$, and the \lstinline[language=xc]|mux| call evaluates to $\infty[]$, leading to:}
\begin{lstlisting}[language=xc,mathescape]
nfold(min, $\infty[\deviceId_1 \mapsto 5, \deviceId_2 \mapsto \infty, \deviceId_3 \mapsto 2]$, $\infty$[])
\end{lstlisting}
\revB{According to rule \ruleNameSize{[A-FOLD]}, we set $\lvalue_0 = \infty$, $\lvalue_1 = \min(\lvalue_0, 5) = 5$, $\lvalue_2 = \lvalue_1$, $\lvalue_3 = \min(\lvalue_2, 2) = 2$, which is the overall result (and is stored in the value-tree as well by rule \ruleNameSize{[A-XC]}).}
\end{exa}


\paragraph*{Network semantics.}

The evolution of a network executing a program is formalised through structures of \emph{events}, atomic computations performed according to the device-level semantics. Such events across space and time (i.e., when and in which devices they happen) form an ``aggregate'' execution in which information is exchanged across events following a \emph{messaging} relationship $\neigh$.
This idea, \revB{informally presented in \Cref{sys-model},} can be formalised through \emph{augmented event structures}, which expands a classical \emph{event structure} \cite{lamport:events} with further information associated to events: device identifiers and sensor status.

\begin{defi}[Augmented Event Structure] \label{def:augmentedES}
	An \emph{augmented event structure} ${\aEventS = \ap{\eventS,\neigh,\devof,\sensof}}$ is a tuple where:
	\begin{itemize}
	\item
	$\eventS$
	is a countable set of \emph{events} $\eventId$,
	\item
	$\neigh \; \subseteq \eventS \times \eventS$ is a \emph{messaging} relation,
	\item
	$\devof : \eventS \to \deviceS$ is a mapping from events to the devices where they happened, and
	\item
	$\sensof: \eventS \to S$ \revA{maps each event $\eventId$} to a \emph{sensors status} $\senstate$ (as in the device-level semantics),
	\end{itemize}
	\revA{
	 such that:
	 \begin{itemize}
		\item $\eventId_1 = \eventId_2$ whenever $\devof (\eventId_1) = \devof(\eventId_2)$ and $\eventId_i \neigh \eventId$ for $i=1,2$  (i.e., all predecessors of an event happened on different devices);
		\item there are no sequences $\eventId_1 \neigh \ldots \neigh \eventId_n \neigh \eventId_1$ (i.e., the $\neigh$ relation is acyclic);
		\item the set $X_\eventId = \bp{\eventId' \in \eventS \mid ~ \eventId' \neigh \ldots \neigh \eventId}$ of events that can reach $\eventId$ in $\neigh$ is finite for all $\eventId$ (i.e., $\neigh$ is well-founded and locally finite).
	 \end{itemize}
	 }
\end{defi}

We say that event $\eventId'$ is a \emph{\supplier} of event $\eventId$ iff $\eventId' \neigh \eventId$. We call \emph{causality relation} the irreflexive partial order $< \; \subseteq \eventS \times \eventS$ obtained as the transitive closure of $\neigh$.

This computational model captures the behaviour of communicating physical devices. The evolution of the state of devices is a chain of $\neigh$-connected events associated to the same device. Communication is captured by $\neigh$-connected events associated to different devices.
Such a computational model
allows us to express programs abstracting from synchronisation,
shared clocks, or regularity and frequency of events.
Informally, following previous work~\cite{Mamei:2004a,DBLP:journals/corr/Lluch-LafuenteL16,viroli:selfstabilisation}, we refer to a  \emph{field of values} as a mapping from devices to values, capturing a global snapshot of the values
produced by the most recent firing of each device.
The evolution over time of a field of values is then a {\it space-time value}, which maps each event
(in an augmented event structure) to a corresponding value.

\begin{defi}[Space-Time Value]\label{def:stvalue}
	Let  $\aEventS = \ap{\eventS,\neigh,\devof,\sensof}$ be an augmented event structure and $\setVS_\type$ be the domain of \excvalues\ of type $\type$.
	A \emph{space-time value} (in $\aEventS$ of type $\type$)  is a function mapping events to \excvalues\ $\dvalue : \eventS \to \setVS_\type$.
\end{defi}

In this model, the evaluation of a program $\emain$ in an augmented event structure $\aEventS$ then produces a space-time value, induced by repeatedly applying the device-level semantics.

\begin{defi}[Program Evaluation on Event structures] \label{def:network:semantics}
  Let $\e$ be an \sysname\ expression of type $\type$ given assumptions $\TtypEnv$.
  Let $\aEventS$ be an augmented event structure whose $\sensof$ includes values of the appropriate type for each sensor and free variable appearing in $\e$.

  Let $\vtree^{\aEventS}_{\e} : \eventS \to \TreeSet$ (where $\TreeSet$ is the set of all value-trees) and $\dvalue^{\aEventS}_{\e} : \eventS \to \setVS_{\type}$ be defined by induction on $\eventId$ in $\eventS$, so that $\eqhlbsopsem{\devof(\eventId)}{\Trees_{\eventId}}{\sensof(\eventId)}{\e}{\dvalue^{\aEventS}_{\e}(\eventId)}{\vtree^{\aEventS}_{\e}(\eventId)}$ where $\Trees_{\eventId} = \{\devof(\eventId') \mapsto \vtree^{\aEventS}_{\e}(\eventId') : \eventId' \neigh \eventId\}$. Then we say that $\dvalue^{\aEventS}_{\e} : \eventS \to \setVS_{\type}$ is the evaluation of expression $\e$ on $\aEventS$.
\end{defi}

\extLmcsEnd

\section{Distributed Collective Processes in XC}\label{sec:contrib}

In this section,
 we present an implementation of \acp{dcp}
 in XC.
First, we characterise the implementation in terms of an abstract notation and formulas on event structures (\Cref{sec:model-es}).
Then, we provide a formalisation of \acp{dcp}
 in terms of the big-step operational semantics for a new XC construct (\Cref{sec:formaliz-spawn}), that can be used to actually \emph{program} \acp{dcp}. \revB{We conclude discussing engineering characteristics of \acp{dcp} (\Cref{sec:disc-abs}) and their practical implementation in the FCPP framework (\Cref{sec:fcppimpl}).}

\subsection{Modelling on event structures}
\label{sec:model-es}

A \emph{\acl{dcp} (\acs{dcp})} $P$ is a computation with given programmed behaviour.
A single DCP can be run in multiple \emph{process instances} $P_i$, each associated to a unique \emph{\ac{pid}} $i$, which we assume also embeds construction parameters for the process instance.
New instances of an aggregate process $P$ are spawned through a \emph{generation field} $G_P$, producing a set of identifiers $G_\revB{P}(\eventId) = \{i\ldots\}$ in each event $\eventId$, of process instances that need to be created in that event $\eventId$ (which we call \emph{initiator} for $P_i$).
For each process instance $P_i$, we use the Boolean predicate $\proc(\eventId)$ to denote whether such instance is being executed at event $\eventId$ (either being initiated by $\eventId$, or through propagation from previous events).
Each process instance $P_i$, if active in an event $\eventId$ (i.e., $\pi_{P_i}(\eventId) = \top$),
locally computes both an \emph{output} $\oproc(\eventId)$ (returned to the process caller)
and a \emph{status} $\sproc(\eventId)$ \revB{according to some given program (possibly involving communication with neighbours)}.
The status
 is an \excvalue\ mapping the device $\devof(\revB{\eventId'})$ of each neighbour event $\eventId' \in \neighset$ to a $\tbool$ value indicating whether that device has to run the same process instance the next round. \revB{The default value of the status determines whether the process instance should be propagated or not to devices that were not neighbours in $\eventId$. Every event that has \emph{at least one} neighbour that propagated the process instance to it will run the process instance as well. More precisely, a} process instance $P_i$ which is active in an event $\eventId$ {\em potentially} propagates the process to any event $\eventId'$ of which $\eventId$ \revB{for which it is} a neighbour ($\eventId \neigh \eventId'$) depending on the value of $\sproc(\eventId)$. In formulas:
\begin{equation*}
	\pi_{P_i}(\eventId) = \begin{cases}
		\top & \text{if } i \in G_P(\eventId),~\text{otherwise} \\
		\top & \text{if } \exists \eventId' \neigh \eventId. ~ \pi_{P_i}(\eventId') ~\wedge~ \sproc(\eventId')(\devof(\eventId)) = \top \\
		\bot & \text{otherwise}.
	\end{cases}
\end{equation*}
\revB{This notion of \acp{dcp}, defined by the abstract model above, is particularly well-suited for being implemented in \sysname, as shown in the following section.}

\begin{exa}[Multi-gradient]\label{ex:multigradient}
As a running example, consider a \emph{multi-gradient} $P^{mg}$: the construction of distinct gradients (cf. \autoref{ex:gradient}) from multiple sources.
A gradient has to be spawned locally in source devices; therefore, the generation logic $G_{P^{mg}}$ could be as follows:
$$
G_{P^{mg}}\revB{(\eventId)} = \begin{dcases}
\{ \revB{\devof(\eventId)} \} & \revB{\textit{isSource}(\eventId)}
\\
\emptyset & \text{otherwise}
\end{dcases}
$$
\revB{where $\textit{isSource}(\eventId)$ is true iff the device $\devof(\eventId)$ is considered a source when event $\eventId$ occurs.}
The output $O_{P_i^{mg}}\revB{(\eventId)}$ is the gradient \revB{as computed in $\eventId$, representing the distance} from source device $i$ \revB{to the current device $\devof(\eventId)$. This can be modelled by the following output computation, analogous to the computation of $g(\eventId)$ as presented in Example \ref{ex:gradient}:
\[
O_{P_i^{mg}}(\eventId) = \begin{dcases}
	\min_{\eventId' \in \neighset(\eventId)\setminus\eprec(\eventId)} \{ O_{P_i^{mg}}(\eventId') +  D_{\eventId}(\devof(\eventId')) \} & \devof(\eventId) \neq i \\
	0 & \devof(\eventId) = i
\end{dcases}
\]}

If each gradient has to be limited in space, say within \revB{a maximum distance of} $\rho$ from the source, then \revB{a possible implementation of} the status logic $s_{P_i^{mg}}$ of a process \revB{instance could be the following:
\[
s_{P_i^{mg}}(\eventId) = \evalue{\left(O_{P_i^{mg}}(\eventId) \le \rho\right)}{i}{\bot}
\]
In this way, $s_{P_i^{mg}}(\eventId)$ is an \excvalue\ that by default correspond to the Boolean value of whether the current gradient value is less than or equal to $\rho$. This \excvalue\ carries a single exception, for the source device $i$, for which the value $\bot$ is used instead, so that the process instance is never propagated to the source. This makes sure that a the process instance can live on the source only as long as the source is in the generation set $G_{P^{mg}}$ (and thus  \textit{isSource} is still true for it). In case $i$ stops being a source, the source itself would immediately leave the process, and for other devices the gradient estimates in $O_{P_i^{mg}}(\eventId)$ would start to rise, until they surpass the threshold $\rho$ effectively terminating  the whole instance.
}
\end{exa}

\subsection{\revB{Formalisation in \sysname}}
\label{sec:formaliz-spawn}

\revB{In XC, the notion of \acp{dcp} is implemented through a single} built-in construct $\spawnxc(P,G_P)$
 that runs independent instances of a \revB{given process} $P$,
 where new instances are locally generated according to \emph{generation field} $G_P$ as explained above.
The output of a $\spawnxc(P,G_P)$ expression in an event $\eventId$
is the set of pairs $\{(i, \oproc(\eventId)), \ldots\}$ for which $\pi_{P_i}(\eventId) = \top$.
%
See \Cref{fig:dcp-example} for an example showing two overlapping process instances.

\begin{figure*}[t]
\centerline{\framebox[\linewidth]{$\begin{array}{l}
	\textbf{Auxiliary definitions:} \\ \quad
	\begin{array}{ccl@{\hspace{6mm}}r @{\hspace{35mm}} l}
		\vtree & \BNFcce & {\mkvt{}{\overline{\vtree}} \; \BNFmid} \; \mkvt{\anyvaluealt}{ {\overline\vtree} } \; \BNFmid \; \eqhl{\envmap{\overline\lvalue}{\overline\vtree}} &   {\footnotesize \mbox{value-tree}}
		\\
	\end{array} \\ \quad
	\begin{array}{@{\hspace{0pt}}l@{\hspace{13pt}}l@{\hspace{13pt}}l}
		\eqhl{\piBof{\lvalue}{\envmap{\overline\lvalue}{\overline\vtree}} = \vtree_i} &
		\eqhl{\textrm{s.t.~} \lvalue_i = \lvalue \textrm{ if it exists}} &
		\eqhl{\textrm{else~} \emptyseq}
		\\
	\end{array} \\
	\hline\\[-10pt]
	\textbf{Auxiliary evaluation rules:} \hspace{40mm} \boxed{\eqhlauxopsem{\deviceId}{\Trees}{\senstate}{\funvalue(\overline\anyvaluealt)}{\anyvaluealt}{\vtree}} \\

	\begin{array}{c}
		\surfaceTyping{A-SPAWN}{\!\!
			\begin{array}{l}
			k_1, \ldots, k_n = \anyvaluealt^k(\deviceId) \cup \{k \text{ for } \deviceId' \in \domof{\Trees}, \envmap{k}{\mkvt{b}{\vtree}} \in \Trees(\deviceId') \text{ with } b(\deviceId) = \truevalue\}
			\\
			\bsopsem{\deviceId}{\piIof{1}{\piBof{k_i}{\Trees}}}{\senstate}{\anyvaluealt^p(k_i)}{\anyvaluealt_i}{\vtree_i} \text{~where~} \anyvaluealt_i = \pairK(\anyvalue_i, b_i) \text{ for } i \in 1, \ldots, n
			\end{array}
		 }{
			\eqhlauxopsem{\deviceId}{\Trees}{\senstate}{\spawnK(\anyvaluealt^p, \anyvaluealt^k)}{\envmap{\overline k}{\overline \anyvaluealt}}{\envmap{\overline k}{\mkvt{\overline{b}}{\overline\vtree}}}
		}
	\end{array}
\end{array}$}}
\caption{Device (big-step) operational semantics of \sysname} \label{fig:spawn:sem}
\end{figure*}

\begin{figure}[t]
\newcommand{\toAbove}{1.5cm}
\newcommand{\toRight}{1cm}
\centering
\adjustbox{trim=0cm 0.4cm 0cm 0cm,max width=\esResizeFactor\columnwidth}{
\begin{tikzpicture}[evlabel/.style={color=red!70!black,font=\scriptsize\bfseries,fill=white, fill opacity=0.8,yshift=-5pt}]
\node[] (origin) at (0,0) {};

\foreach \dev/\evs/\offx/\freq in {%
1/3/ -0.8 / 3.5,%
2/5/ -0.2 / 2.3,%
3/4/ 0.3 / 2.6,%
4/6/ -0.2 / 1.8,%
5/3/ -0.8 / 3.8}
  \foreach \ev in {1,...,\evs}
        \node[present,font=\Large] (d\dev\ev) [above right = \dev*\toAbove and \ev*\toRight*\freq,xshift=-1cm+\offx*2cm,yshift=-1cm] {$\eventId_\ev^\dev$}; 

\makeatletter
\long\def\ifnodedefined#1#2#3{%
    \@ifundefined{pgf@sh@ns@#1}{#3}{#2}%
}
\makeatother
\foreach \ref in {d3e2} {
\foreach \a in {1,...,5} {
 \foreach \ea in {1,...,10} {
  \foreach \b in {1,...,5} {
  \foreach \eb in {1,...,10} {
  \newtoggle{closer}
   \ifnodedefined{d\a\ea}{
   	  \ifnodedefined{d\b\eb}{
   	  	\ifthenelse{\NOT \equal{\a}{\b} \OR \NOT \equal{\ea}{\eb} }{ 
		  \draw[->]
     		let \p1=(d\a\ea), \p2=(d\b\eb),
       		  \n3={\y2-\y1},
     		  \n2={\x2-\x1},
     		  \n1={veclen(\x2-\x1,\y2-\y1)} in
     		{\ifdim \n1 < 3.2cm
     		 \ifdim \n2 > 0.8cm
     		 (d\a\ea) -- (d\b\eb)
     		 \fi
     		 \fi};
     	}{}
   	  }{}
   }{}
  }
}}}}

\newcommand{\nlabel}[3]{
  \foreach \lbl [count=\lbli] in {#2}
    \node [evlabel,#3] (ld#1\lbli) [above=0cm of d#1\lbli] {$\lbl$};
}

\foreach \a in {1,...,5} {
 \foreach \ea in {1,...,10} {
 \pgfmathtruncatemacro\next{\ea+1}
 \ifnodedefined{d\a\next}{
 	\draw[->] (d\a\ea) -- (d\a\next);
}}}

\begin{scope}[on background layer]
    \filldraw[\colorRegionA, line join=round, line width=1cm, opacity=0.8] plot coordinates{(d21.west) (d31) (d43) (d46) (d34.north east) (d25.east) (d13) (d12) (d21.west)}--cycle;
    \filldraw[\colorRegionB, line join=round, line width=1cm, opacity=0.6] plot coordinates{(d51.north west) (d31) (d22) (d32)  (d44.east) (d43)  (d51.north east)}--cycle;

\end{scope}

\draw[->] (0,0) -- node[xshift=-1cm]{Devices} (0,7);
\draw[->](0,0) -- node[yshift=-0.6cm]{Time} (10,0);

\node (gen1) [above=0.25cm of d51] {$G(\eventId_1^5)=\{5\}$};
\node (gen2) [below=-0.2cm of d21] {$G(\eventId_1^2)=\{2\}$};
\node (endsource1) [above=0cm of d52] {$\mathit{source}=\bot$};

\nlabel{5}{{[5 \mapsto{} 0]} 
}{}
\node [evlabel, xshift=-1pt] (ld42) [above=0cm of d42] {$[5 \mapsto{} 1]$};   
\node [evlabel, xshift=-4.5pt] (ld43) [above=0cm of d43] {$[2 \mapsto{} 2, 5 \mapsto{} 1]$};
\node [evlabel, xshift=4.5pt] (ld44) [above=0cm of d44] {$[2 \mapsto{} 2, 5 \mapsto{} 3]$};
\node [evlabel, xshift=1pt] (ld45) [above=0cm of d45] {$[2 \mapsto{} 2]$};
\node [evlabel] (ld46) [above=0cm of d46] {$[2 \mapsto{} 2]$};
\nlabel{3}{{[2\mapsto{}1, 5 \mapsto 1]},
 {[2 \mapsto 1, 5 \mapsto 2]},
 {[2 \mapsto 1]},
 {[2 \mapsto 1]}
 }{}
\nlabel{2}{{[2 \mapsto{} 0]}, {[2 \mapsto{} 0, 5 \mapsto 2]}, {[2 \mapsto{} 0]} ,{[2 \mapsto{} 0]}, {[2 \mapsto{} 0]}}{}
\node [evlabel] (ld12) [above=0cm of d12] {$[2 \mapsto{} 1]$};
\node [evlabel] (ld13) [above=0cm of d13] {$[2 \mapsto{} 1]$};

\end{tikzpicture}
}
\caption{\extLmcs{\revB{This diagram shows the dynamics of a multi-gradient (cf. \autoref{ex:multigradient}) with devices 2 and 5 as sources, where the former is a source for the entire execution shown, and the latter only in its first event, $\eventId_1^5$.
Two process instances 
   $P_{5}$ (yellow) and $P_{2}$ (blue),
   each one carrying a gradient process,
   are generated from events $\eventId_1^5$ and $\eventId_1^2$, respectively.
Suppose both processes are instantiated to propagate within a maximum number $\rho=1$ of hops from the initiator device. 
The two process instances \emph{overlap} in various events of devices 2 ( $\eventId_2^2$), 3 ($\eventId_1^3$,  $\eventId_2^3$), and 4 ($\eventId_3^4$, $\eventId_4^4$, )---where both instances are run at the same time in non-deterministic order.
Since it is  assumed that the device 5 is a source only at $\eventId_1^5$; 
then, devices 4, 3, and 2 start bubbling the gradient value up, until exiting the process. 
Recall that any event propagates the process if and only if its gradient value is less than or equal $\rho=1$, that process instances are independent, and that the gradient values adds a unit of distance to the minimum across neighbours events (excluding self).
}
}
}
\label{fig:dcp-example}
\end{figure}

The $\spawnxc$ construct \revB{corresponds to the mathematical definition given} in the previous section (\revA{\Cref{sec:model-es}}), and embeds naturally \revA{in XC as a built-in function}. 
As a built-in in \sysname, $\spawnK$ assumes the same type as the classical spawn construct in \ac{fc}: $\forall \tvar_k,\tvar_v. ((\tvar_k) \to \tpair[\tvar_v, \tbool], \tset[\tvar_k]) \to \tmap[\tvar_k,\tvar_v]$. However, in \sysname\ every type allows \excvalues, which translates into practical differences.

Figure \ref{fig:spawn:sem} presents the semantics of the spawn built-in, relative to the \sysname\ semantics \revB{presented in \Cref{ssec:opsem}}.
In order to introduce the $\spawnK$ construct, it is necessary to extend the auxiliary definition of value-trees (highlighted in grey), to also allow for maps from local literals $\lvalue$ (identifiers of the running processes) to their corresponding value-trees. Then, rule \ruleNameSize{[A-SPAWN]} can be written by naturally porting the similar rule in \cite{casadei2019aggregate}, while using the fact that the Boolean returned by the process is an \excvalue, and thus can be different for different neighbours.
\revA{In this rule, a list of \emph{process keys} $\overline k$ is computed by adjoining
\emph{(i)} the keys $\anyvaluealt^k(\deviceId)$ currently present in the second argument $\anyvaluealt^k$ of $\spawnK$ for the current device $\deviceId$;
\emph{(ii)} the keys that any neighbour $\deviceId'$ broadcast in their last message $\Trees(\deviceId')$, provided that the corresponding Boolean value $b$ returned was true for the current device $b(\deviceId) = \truevalue$ (thus, demanding process expansion to $\deviceId$).
To realise ``multiple alignment'', for each key $k_i$, the process $\anyvaluealt^p$ is applied to $k_i$ with respect to the part of the value-tree environment $\piIof{1}{\piBof{k_i}{\Trees}}$ that corresponds to key $k_i$, producing $\anyvaluealt_i;\vtree_i$ as a result. Finally, the construct concludes returning the maps $\envmap{\overline k}{\overline \anyvaluealt};\envmap{\overline k}{\overline \vtree}$ mapping process keys to their evaluation result.}

\begin{exa}[\revB{Multi-gradient Implementation}]\label{ex:multigradient:impl}
\revB{As an example, the multi-gradient introduced in Example \ref{ex:multigradient} can be translated in \sysname\ obtaining the following code:}

\begin{lstlisting}[language=xc]
def nbr(v) {
  exchange( v, (n) => (n,v) )
}
def multi_gradient(isSource, theta) {
  val gen_set = if (isSource) {set(uid())} else {set()};
  spawn((i) => {
    val output = gradient(uid() == i);
    val status = output <= theta && nbr(uid()) != i;
    (output, status)
  }, gen_set);
}
\end{lstlisting}

\revB{
The multi-gradient function receives as a parameter the Boolean value \emph{isSource} identifying source devices and the parameter $\rho$ regulating the size of the areas spanned by each process. The generation set \lstinline[language=xc]|gen_set| is computed following exactly the definition of $G_{P^{mg}}$, and similarly for \lstinline[language=xc]|output| with $O_{P_i^{mg}}$, by relying on the \sysname\ implementation of the gradient discussed in Example \ref{ex:gradient}.
In order to define \lstinline[language=xc]|status| as in $s_{P_i^{mg}}$, the translation starts from the same default value \lstinline[language=xc]|output <= theta| of $s_{P_i^{mg}}$, and adds the exception for $i$ by conjunction with an \excvalue\ that is false only for $i$: \lstinline[language=xc]|nbr(uid()) != i|. 

In the evaluation of the \lstinline[language=xc]|multi_gradient| function, the application of \lstinline[language=xc]|spawn| is performed according to rule \ruleNameSize{[A-SPAWN]}.
First, the set of currently active process keys is retrieved from the value-tree environment $\Trees$. For instance, in the example described in \Cref{fig:dcp-example}, for event $\eventId_2^3$, we would have $\textit{isSource} = \bot$,  $\rho = 1$, and $\Trees$ equal to:
\begin{align*}
	\deviceId_2 &\mapsto \cp{\deviceId_2 \mapsto \mkvt{\top[\deviceId_2 \mapsto \bot]}{\vtree^2_2}, \deviceId_5 \mapsto \mkvt{\bot[]}{\vtree^5_2}}, \\
	\deviceId_3 &\mapsto \cp{\deviceId_2 \mapsto \mkvt{\top[\deviceId_2 \mapsto \bot]}{\vtree^2_3}, \deviceId_5 \mapsto \mkvt{\top[\deviceId_5 \mapsto \bot]}{\vtree^5_3}}, \\
	\deviceId_4 &\mapsto \cp{\deviceId_5 \mapsto \mkvt{\top[\deviceId_5 \mapsto \top[\deviceId_5 \mapsto \bot]]}{\vtree^5_4}}
\end{align*}
In this case, the generation set is empty ($\anyvaluealt^k(\deviceId)$ in rule \ruleNameSize{[A-SPAWN]}). 
Thus, the set of active process keys is computed only by considering, for every $\deviceId'$ in $\domof{\Trees}$ (that is, $\deviceId_2$, $\deviceId_3$, $\deviceId_4$) their corresponding keys \emph{only if} associated with a true Boolean for the current device $\deviceId_3$.
The key $\deviceId_5$ coming from the neighbour $\deviceId$ is discarded, since its corresponding status is $\bot[]$ not true for $\deviceId_3$. All other keys have a corresponding status that is true for $\deviceId_3$, so the keys computed are $k_1 = \deviceId_2, k_2 = \deviceId_5$.

Then, for $i= 1,2$ the process body is executed with respect to the environment corresponding to the same process execution in neighbours. For example, for $k_1 = \deviceId_2$:
\[
	\pi^{\deviceId_2}(\Trees) = \deviceId_2 \mapsto \mkvt{\top[\deviceId_2 \mapsto \bot]}{\vtree^2_2}, \deviceId_3 \mapsto \mkvt{\top[\deviceId_2 \mapsto \bot]}{\vtree^2_3}, \quad
	\pi_1(\pi^{\deviceId_2}(\Trees)) = \deviceId_2 \mapsto \vtree^2_2, \deviceId_3 \mapsto \vtree^2_3
\]
and the body of the process is executed with argument $\deviceId_2$ w.r.t. this environment. The execution will result in the pair $(1[], \top[\deviceId_2 \mapsto \bot])$ which will contribute by adding $\deviceId_2 \mapsto 1[]$ to the overall result and $\deviceId_2 \mapsto \mkvt{\top[\deviceId_2 \mapsto \bot]}{\ldots}$ to the produced value-tree.
}
\end{exa}

\subsection{The \ac{dcp} abstraction}
\label{sec:disc-abs}

The crucial problem that we investigate in this paper
 revolves around the definition
 of collaborative activities
 carried out by \emph{dynamic} collections of devices (a.k.a. \emph{ensembles}~\cite{DBLP:journals/taas/NicolaLPT14}).
We call these \emph{\acp{dcp}} since they are defined by a common control program that regulates the behaviour of a largely homogeneous set of devices.

In particular, a device may participate \emph{concurrently} to multiple collective processes, or to multiple ensembles; \revB{and} in any single round,
 a device executes the computation associated
 to \emph{all} the currently joined \acp{dcp}.
\revA{
Participation to multiple \acp{dcp} \revB{may thus have consequences on} local resource usage (cf. resource-constrained devices). Therefore, real-world implementations have also to \revB{take performance issues into account. Since there is no universal solution for managing resource constraints in devices participating to multiple DCPs, much of the responsibility on handling these situations lies with the programmer. Some general strategies may be applicable, such as:
\emph{(i)} tuning the scheduling policy according to the number of processes joined by a device, using extensions of the calculus that have been previously studied \cite{DBLP:journals/lmcs/PianiniCVMZ21} and are already available in current implementations (including FCPP \cite{a:fcpp});
\emph{(ii)} computing a status from the resource availability perceived through a sensor or runtime checks, using it to prevent process propagation to nodes that are already on a high load;
\emph{(iii)} designing \acp{dcp} in order to minimise resource consumption, both through low-level optimisations and by reducing the area spanned by a single \acp{dcp} to the smallest possible.
Further details on these strategies go beyond the scope of this paper.
}
}

We define as the \emph{domain} of a \ac{dcp}
 the set of the nodes that are currently running it.
We define as the \emph{shape} of a \ac{dcp}
 the spatiotemporal region that
 is identified by the spatiotemporal locations of the nodes
 belonging to the domain of the \ac{dcp}.
Often, \acp{dcp} are \emph{transient}, i.e.,
 they have a limited lifetime:
 they start to exist at some time,
 and they dissolve once no more nodes run them.

In this work,
 we are mainly concerned with
 studying how to create and manipulate these \acp{dcp}.
The supporting formal framework
 and implementation is described in \Cref{sec:contrib}.
\revA{
In summary, the developer has the following mechanisms for defining systems of \acp{dcp}:
\begin{itemize}
\item \emph{generation logic}: the need for collective activities can be encoded in a rule for generating new instances of \acp{dcp};
\item \emph{identity logic}: the logic used to identify \acp{dcp} can be used to distinguish between them and hence to regulate their domains (e.g., for controlling the granularity of teams);
\item \emph{internal logic}: this logic defines a collective computation (scoped within the domain of a single \ac{dcp}) promoting decentralised decision-making, e.g., in terms of typical self-organisation patterns (collection, propagation, leader election, evaporation, etc.);
\item \emph{shape control logic}: it is possible to specify rules for the local expansion of the domains of \acp{dcp} (e.g., to gather more participants), typically also leveraging results from the internal computation itself---the XC implementation provides an especially flexible way to specify this, as different neighbours can receive different information;
\item \emph{termination logic}: this logic, strictly related to shape control, enables to specify how individual agents may leave a \ac{dcp} instance as well as how an entire \ac{dcp} may be terminated;
\item \emph{input logic}: existing \acp{dcp} may also be controlled be specifying ``external inputs'' provided as explicit arguments or closed over a lambda closure---an example is \emph{meta-control logic}, based on inspecting the (results of) multiple \acp{dcp} to take decisions about their evolution.
\end{itemize}
}

\extLmcs{
\subsection{Implementation in FCPP}\label{sec:fcppimpl}
We now briefly describe the implementation of $\spawnxc$ in the FCPP C++ framework \cite{a:fcpp}, since it is used in the code examples in the next sections.
As typical of the style adopted in FCPP, $\spawnxc$ is defined as a number of overloads of a template function.
Here we focus on the overload that allows the process function to return a pair containing its output value and the status {\excvalue} which specifies for each neighbour $\deviceId'$ (possibly also the device itself) whether the instance should be propagated to $\deviceId'$ or not.
The signature of the template function is the following:
}

\begin{lstlisting}[language=C++,
                   basicstyle=\footnotesize\ttfamily\lst@ifdisplaystyle\footnotesize\fi,
                   keywordstyle=\color{blue},
                   frame=single,
                   basewidth=0.5em]
  template<typename node_t, typename G, typename S, typename... Ts,
    typename K = typename std::decay_t<S>::value_type,
    typename T = std::decay_t<std::result_of_t<G(K const&, Ts const&...)>>,
    typename R = std::decay_t<tuple_element_t<0,T>>,
    typename B = std::decay_t<tuple_element_t<1,T>>>
  std::enable_if_t< std::is_same< B, nvalue<bool> >::value,
                    std::unordered_map< K, R, common::hash<K> > >
  spawn(node_t& node, trace_t call_point, G&& process, S&& key_set, Ts const&... xs)
\end{lstlisting}

\extLmcs{
The first line of the signature specifies the type parameters of the function: \texttt{node\_t} (device object), \texttt{G} (process function), \texttt{S} (\revB{container of} keys of process instances), and the variadic types \texttt{Ts} of other arguments to the process function.
The \revB{subsequent} lines declare further types \revB{deduced} from the parameters: \texttt{K} (\revB{type of the keys in the container} \texttt{S}), \texttt{T} (full return type of the process function), \texttt{R} (output value returned by the process function), and \texttt{B} (propagation value returned by the process function).

Then, a further constraint is declared on these types through \texttt{std::enable\_if\_t}. In particular the \texttt{is\_same} constraint requires that type \texttt{B} should be an {\excvalue} of Booleans, i.e., that this particular overload of $\spawnxc$ can propagate process instances differently to different neighbours.

Despite the complexity of the signature, due to the flexibility offered by templates, a call to the $\spawnxc$ in FCPP can be as simple as:
}

\begin{lstlisting}[language=C++,
  basicstyle=\footnotesize\ttfamily\lst@ifdisplaystyle\footnotesize\fi,
  keywordstyle=\color{blue},
  morekeywords={CALL},
  frame=single,
  basewidth=0.5em]
  auto r = spawn(CALL, proc_func, new_instances, xarg1, ..., xargk)
\end{lstlisting}

\extLmcs{
where \texttt{CALL} is a FCPP macro that automatically provides arguments \texttt{node} and \texttt{call\_point}, \texttt{proc\_func} is a lambda or pointer to the process function, \texttt{new\_instances} is a collection of process instance keys to be spawned, and the \texttt{xarg}s are the optional extra arguments to \texttt{proc\_func}.

In FCPP, the code for the process function of the multi-gradient \ac{dcp} \revB{presented} in Example \ref{ex:multigradient:impl} is \revB{translated} as follows:
}

\begin{lstlisting}[language=fcpp]
FUN real_t gradient(ARGS, bool isSource) { CODE
  return exchange(CALL, INF, [&](nvalue<real_t> n){
    return min_hood(CALL, n + node.nbr_dist(), mux(isSource, 0, INF));
  });
}

FUN std::unordered_map<device_t, real_t>
multi_gradient(ARGS, bool isSource, real_t theta) { CODE
  common::option<device_t> gen_set;
  if (isSource) gen_set = node.uid;
  return spawn(CALL, [&](device_t i) {
    real_t output = gradient(CALL, uid() == i);
    nvalue<bool> status = output <= theta && nbr(CALL, node.uid) != i;
    return make_tuple(output, status);
  }, gen_set);
}
\end{lstlisting}

\revB{As apparent, the code has the same structure as the code in Example \ref{ex:multigradient:impl}, except for the addition of explicit types to variables according to the standard C++ syntax, and the addition of the FCPP-specific macros for passing the context through function calls. Recall that, in C++, syntax \lstinline[language=fcpp]|[&](args){body}| is for anonymous functions.}

\section{\extLmcs{Case Studies}}
\label{sec:eval}
\label{ssec:poc}

\extLmcs{
In this section, we present in detail two case studies to illustrate the application of the techniques described in the paper to realistic problems.
\revB{
We cover first the case study on multi-hop message propagation (\Cref{eval:msg-propagation})
 and then the case study on spatial monitoring (\Cref{sec:eval:pastctl}): for both of them, we discuss the implementation, the experimental setup (scenario, parameters, metrics),
 and finally the results.
Both t}he use cases have been implemented  exploiting the FCPP simulator \cite{a:fcpp,DBLP:conf/coordination/AudritoRT22}, which has been extended to support the XC and, in particular, the $\spawnxc$ built-in construct described above.
\revB{
  The experimental framework is publicly available\footnote{\url{https://github.com/fcpp-experiments/xc-processes}\label{footnote:eval-repo}}, with build infrastructure and instructions, for inspection and reproducibility.
}
}

\subsection{\extLmcs{Message Propagation}}\label{eval:msg-propagation}

We consider a network of devices where, at some point in time, a source device $\deviceId_{FROM}$ sends a message through a \ac{dcp} to reach a destination device $\deviceId_{TO}$.
\revB{
The goal is to show that the \ac{dcp} mechanism
  \emph{works} for the specific use case,
  that multiple strategies exhibiting different dynamics are possible,
  and that the \ac{xc} implementation
  can unlock performance benefits with respect to field calculus-based implementations.
}

\subsubsection{Implementation}
\extLmcs{We have experimented two different ways of propagation of the \ac{dcp}.}
The first one is a \emph{spherical} propagation, where the message originating in $\deviceId_{FROM}$ spreads radially in 3D trying to reach $\deviceId_{TO}$. The process function executed by each process node implements the following logic:
\revB{
\begin{enumerate}
\item if the self device $\deviceId$ is $\deviceId_{TO}$ just return a $\bot$ \excvalue\ $F[]$ (no propagation, since destination has been reached);

\item if it is the first round that $\deviceId$ executes the process, and the neighbours it knows (i.e., that have propagated the process to $\deviceId$) are $\deviceId_1, \ldots, \deviceId_k$, it propagates it to itself and to its {\em new neighbours} (that are not yet in the process) by returning an \excvalue\ $T[\deviceId\!\mapsto\!T, \deviceId_1\!\mapsto\!F, \ldots]$;

\item if it is the second round that $\deviceId$ executes the process, it propagates the process only to itself

\item finally, at the third round $\deviceId$ exits itself the process by returning a $\bot$ \excvalue\ $F[]$.
\end{enumerate}
}

The snippet of FCPP code in \Cref{fig:msgpropcode} shows the call to $\spawnxc$ that implements the \ac{dcp} described above.

\begin{figure}
\begin{lstlisting}[language=fcpp,mathescape]
message_log_type r = spawn(CALL, [&](message const& m) {
  nvalue<bool> nv_status = false; // nvalue $\bot[]$
  bool target_node = (m.to == node.uid);
  int round = counter(CALL); // just counts the number of rounds run by the device

  if (not target_node and round <= 2) {
    nv_status = mod_self(CALL, nv_status, true);
    nv_status = mod_other(CALL, nv_status, round==1);
  }

  return make_tuple(node.current_time(), nv_status);
}, key_set);
\end{lstlisting}
\caption{FCPP code corresponding to the logic of the \ac{dcp} for the message propagation scenario.}
\label{fig:msgpropcode}
\end{figure}

\extLmcs{
  The process identifiers of the \ac{dcp} instances are \texttt{message} objects, which specify the source device \texttt{from}, the target device \texttt{to}, and the payload.
  Note that, in the body of the process function: flag \texttt{target\_node} is $\top$ only on device $\deviceId_{TO}$; \texttt{nv\_status} is the {\excvalue} that determines process propagation; and a call to \texttt{mod\_self} (resp. \texttt{mod\_other}) sets the value for the current device (resp. the default value, excluding current neighbours) in an {\excvalue}.
}

  \extLmcs{The second strategy is a propagation on a path along a \emph{tree}, where the message originating in $\deviceId_{FROM}$ spreads along a spanning tree of the network of devices to reach $\deviceId_{TO}$.
    It is worth pointing out that the spanning tree itself is computed in a fully distributed way by the \ac{xc} program running on each device. Here, we are just interested in the fact that during such computation each node $\deviceId$ holds three {\excvalues} that are used by {\spawnxc} (see the snippet below):
\revB{
  \begin{itemize}
  \item \texttt{nv\_id}, containing the ids of all the neighbours $\deviceId'$ of $\deviceId$ (including $\deviceId$ itself);
  \item \texttt{nv\_parents}, containing the ids of the parents of all the neighbours $\deviceId'$ of $\deviceId$ in the spanning tree;
  \item \texttt{nv\_below}, containing, for each neighbour $\deviceId'$, the set of ids of the nodes below $\deviceId'$ in the spanning tree.
  \end{itemize}
}

The process function executed by each process node implements the following logic:
\revB{
  \begin{enumerate}
  \item as in the spherical propagation, if the self device $\deviceId$ is $\deviceId_{TO}$ just return a $\bot$ \excvalue\ $F[]$ (no propagation, since destination has been reached);

  \item if it is the first or second round that $\deviceId$ executes the process, two {\excvalues} are computed: \texttt{nv\_up} (resp. \texttt{nv\_down}) where each neighbour is associated with $\top$ iff it is the parent of $\deviceId$ in the path from $\deviceId_{FROM}$ to the root of the spanning tree (resp. it is the child of $\deviceId$ in the path from the root of the spanning tree to $\deviceId_{TO}$);
    taking the logical \emph{or} of these {\excvalues} yields an {\excvalue} \texttt{nv\_status} which associates a $\top$ value only to the {\em next} device $\deviceId'$ on the path from $\deviceId_{FROM}$ to $\deviceId_{TO}$ along the spanning tree

    \begin{enumerate}
    \item if it is the first round, $\deviceId$ is added to \texttt{nv\_status} to propagate the \ac{dcp} just to devices $\deviceId$ and $\deviceId'$;
    \item if it is the second round, $\deviceId$ is no longer added to \texttt{nv\_status}, and the \ac{dcp} only propagates to $\deviceId'$;
    \end{enumerate}
  \end{enumerate}
}
  The code in \Cref{fig:src-tree} shows the call to $\spawnxc$ that implements the \ac{dcp} described above.
  }
  \begin{figure}[t!]
  \begin{lstlisting}[language=fcpp]
    message_log_type r = spawn(CALL, [&](message const& m) {
      nvalue<bool> nv_status = false;
      bool target_node = (m.to == node.uid);
      int round = counter(CALL);

      if (not target_node and round <= 2) {
        nvalue<bool> nv_up  = map_hood([&](device_t d) {return d == parent;}, nv_id);
        nvalue<bool> nv_down = map_hood([&](device_t d) {return d == node.uid;}, nv_parent)
                     and map_hood([&] (set_t s) {return (s.count(m.to) > 0);}, nv_below);

        nv_status = nv_up or nv_down;
        nv_status = mod_self(CALL, nv_status, round == 1);
      }

      return make_tuple(node.current_time(), nv_status);
    }, key_set);
  \end{lstlisting}
  \caption{\extLmcs{FCPP code for the message propagation along a spanning tree.\label{fig:src-tree}}}
\end{figure}
\revB{
  \extLmcs{
    The part that differs from the previous case is the code executed by $\deviceId$ at the first and second round, in case $\deviceId$ is not the target of the message.
    The two {\excvalues} \texttt{nv\_up} and \texttt{nv\_down} explained above are computed by applying the \texttt{map\_hood} builtin function to the {\excvalues} \texttt{nv\_id}, \texttt{nv\_parent}, and \texttt{nv\_below} resulting from the computation of the spanning tree. As suggested by the name, \texttt{map\_hood} transforms an {\excvalue} into another {\excvalue} (possibly of different type) by applying a given function to each element.
  }
}
\subsubsection{Setup: Scenario, Parameters, Metrics}\label{subsec:msg-propagation-setup}
\revB{
For simplicity, we assume devices to be stationary in the simulation (\emph{speed} equal to $0$), thus inducing a fixed network topology (although the proof-of-concept program could be run with dynamic topologies as well).
We consider a network of $578$ devices into a volume of approximately $1350 m \times 1350 m \times 20 m$, with a connection range of $100 m$; so that the network spans about 20 \emph{hops} and devices have an average neighbourhood size (\emph{dens}) of 10. Devices periodically compute the program about every second for $50$ seconds of simulated time, with a standard deviation on time intervals (\emph{tvar)} of $5\%$.
We measure the following metrics over the time of execution of the simulations:
\begin{enumerate}
\item the average number of active processes {\em aproc} within the network of devices
\item the total number of messages {\em dcount} correctly delivered to their destinations
\end{enumerate}
The metrics shown in the figures below are obtained by averaging on 100 executions.
}

  \begin{figure*}[h!]
    \centering
    \includegraphics[width=\figResizeFactor\linewidth{}]{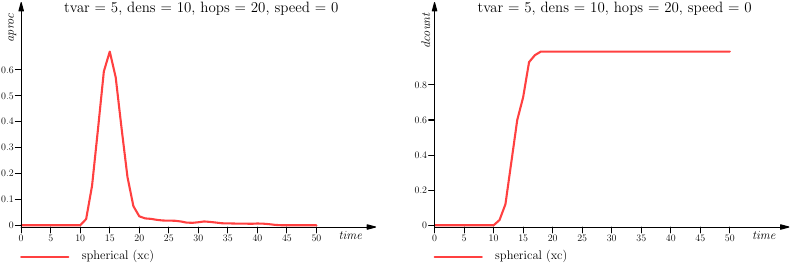}
    \caption{Average number of active processes over time for the single-process use case \revB{and spherical message propagation}.} \label{fig:sphere}
  \end{figure*}
  \begin{figure*}[h!]
    \centering
    \includegraphics[width=\figResizeFactor\linewidth{}]{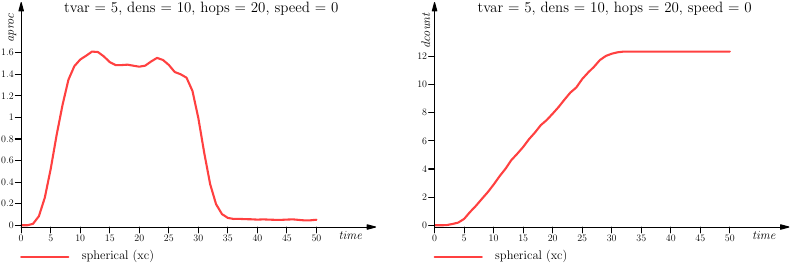}
    \caption{Average number of active processes over time for the multi-process use case \revB{and spherical message propagation}.} \label{fig:sphere-multi}
  \end{figure*}

  \subsubsection{Results}
\revB{
We exploit FCPP to simulate a first use case where only one process is generated. The simulation shows that the process propagates \emph{as a wave} starting from $\deviceId_{FROM}$ outwards. Immediately after the wave front goes beyond a device, the device itself exits the process thus releasing potentially precious resources for other computations.
\Cref{fig:sphere} (left) shows the {\em aproc} metric.
For the first $10$ sec, the average is $0$, since no process has been created yet.
After the process is created by $\deviceId_{FROM}$, it propagates until it reaches its destination, and then quickly vanishes.
Moreover, from \Cref{fig:sphere} (right) showing the {\em dcount} metric, we can see that the message is always delivered.
}

\revB{
  In a second configuration of the use case, we let $10$ different devices generate a new process at each round with probability $5\%$, in the time interval $[1,25]$.
  \Cref{fig:sphere-multi} (left) shows that {\em aproc} grows from $0$ (at time $0$) up to about $1.6$ (approximately at times $10$ and $25$), and then quickly drops again down to $0$. More than $12$ processes are generated during the use case, see the {\em dcount} metric in \Cref{fig:sphere-multi} (right); however, the average number of active processes {\em aproc} is kept low by the fact that the nodes exploit $\spawnxc$ to immediately exit processes after entering and propagating them.
}

\begin{figure*}[h!]
  \centering
  \includegraphics[width=\figResizeFactor\linewidth{}]{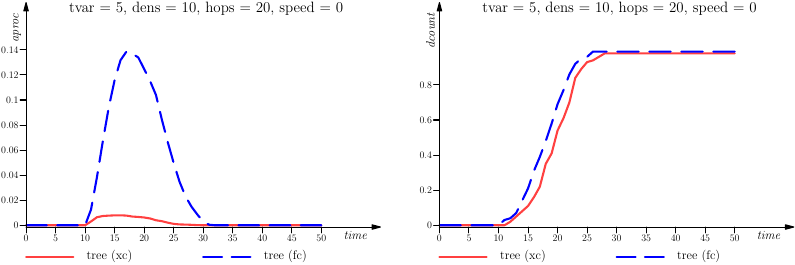}
  \caption{\revB{Average number of active processes over time and total delivered messages for the tree propagation use case. Comparison between $\spawnfc$ and $\spawnxc$.}} \label{fig:tree}
\end{figure*}

\revB{
\extLmcs{
  \Cref{fig:tree} shows the {\em aproc} and {\em dcount} metrics for the tree propagation use case, when only one process is created.
  The red graphics were generated using $\spawnxc$. It is clear that the {\em aproc}, which tops at about $0.01$, is much smaller than in the case of spherical propagation, which tops at more than $0.7$ (\Cref{fig:sphere}).
  In a real scenario, a tradeoff between the reliability and efficiency offered by the two strategies should be considered.

  We have also computed the metrics using $\spawnfc$, i.e., the FC-based spawn operator also available in FCPP (blue dashed graphics). In order to compare as fairly as possible between \ac{xc} and \ac{fc}, we have adopted a wave-like termination for the processes generated by $\spawnfc$ \cite{audrito:fcgs:24}.  Again, we note a more than 10-fold reduced maximum resource-consumption when using $\spawnxc$, since {\em aproc} tops at $0.14$ for $\spawnfc$.
  Therefore, the ability of \ac{xc} to return different states for different neighbours pays off in this scenario.
}
}

\subsection{\extLmcs{Monitoring through Replication}}\label{sec:eval:pastctl}

The second scenario of our evaluation has the main purpose of illustrating another archetypal usage of distributed collective processes, that of \emph{replicated algorithms} \cite{DBLP:conf/coordination/PianiniBV16}, applying it in the context of \emph{distributed monitoring} \cite{DBLP:journals/jss/AudritoDSTV22,adsv:spatialrv}.

\subsubsection{Implementation}
In this case study, we consider a network of devices that may experience some critical situation, modelled by a predicate \emph{critic}. In order to monitor the occurrence of such situations, distributed monitors need to be put in place. Previous works introduce two different abstract logics that can conveniently express such monitors, while granting an automatic translation into an aggregate program realising them: Past-CTL \revB{\cite{past:logics,gmr:past-ltl-syntax,DBLP:journals/jss/AudritoDSTV22}} and the Spatial Logic of Closure Spaces (SLCS) \revB{\cite{DBLP:journals/corr/CianciaLLM16,DBLP:conf/ifipTCS/CianciaLLM14,adsv:spatialrv}}.
The former allows to express properties about the past, such as $\EP \phi$, to be read as \emph{``$\phi$ has been true at some point in the past for some device in the network''}. However, Past-CTL does not capture a notion of what is happening through space at the \emph{present time}. This is possible with the latter logic SLCS, which comprises predicates such as $\SF \phi$, to be read as \emph{``$\phi$ is currently true in some point in space''}.

\begin{figure}[!t]
\centering\revB{
\centerline{\framebox[\linewidth]{$
	\begin{array}{l@{\,}c@{\,}l@{\hspace{-2mm}}r}
	    \formula & \BNFcce & \bot \BNFmid \top \BNFmid \prop \BNFmid (\neg \formula) \BNFmid (\formula \!\wedge\! \formula) \BNFmid (\formula \!\vee\! \formula) \BNFmid (\formula \!\Rightarrow\! \formula) \BNFmid (\formula \!\Leftrightarrow\! \formula)
	    & {\footnotesize \mbox{logical}}
		\\[3pt]
		& & \BNFmid (\DF \formula) \BNFmid (\AF \formula) \BNFmid (\EF \formula) \BNFmid (\DG \formula) \BNFmid (\AG \formula) \BNFmid (\EG \formula)
		& {\footnotesize \mbox{temporal}}
		\\[3pt]
		& & \BNFmid (\DX \formula) \BNFmid (\AX \formula) \BNFmid (\EX \formula) \BNFmid (\formula \DU \formula) \BNFmid (\formula \AU \formula) \BNFmid (\formula \EU \formula)
		\\[6pt]
		\hline
		\\[-6pt]
	    \formula & \BNFcce & \bot \BNFmid \top \BNFmid \prop \BNFmid (\neg \formula) \BNFmid (\formula \!\wedge\! \formula) \BNFmid (\formula \!\vee\! \formula) \BNFmid (\formula \!\Rightarrow\! \formula) \BNFmid (\formula \!\Leftrightarrow\! \formula)
	    & {\footnotesize \mbox{logical}}
		\\[3pt]
		& & \BNFmid (\SI \formula) \BNFmid (\SC \formula) \BNFmid (\SB \formula) \BNFmid (\SBI \formula) \BNFmid (\SBC \formula)
		& {\footnotesize \mbox{spatial}}
		\\[3pt]
		& & \BNFmid (\formula \SR \formula) \BNFmid (\formula \ST \formula) \BNFmid (\formula \SU \formula) \BNFmid (\SG \formula) \BNFmid (\SF \formula) \\[3pt]
	\end{array}
$}
}}
\caption{\revB{Syntax of past-CTL \arevC{(top)} and SLCS \arevC{(bottom)}.}} \label{fig:syntax}
\end{figure}

\revB{
Figure \ref{fig:syntax} illustrates the syntax of the past-CTL and SLCS logics \revD{(as presented in \cite{adsv:spatialrv})}, which build on atomic propositions $\prop$ to represent observables, using standard logical operators alongside additional temporal or spatial modalities.

\arevC{In this framework, past-CTL formulas $\formula$ can be interpreted on augmented event structures $\aEventS$ (\Cref{fig:event-structure-example}), \arevC{yielding a truth value in each event $\eventId$.}} Temporal modalities mirror traditional CTL but with two key distinctions:
\begin{itemize}
	\item Temporal operators apply to \arevC{actual \emph{(message) paths} in $\aEventS$, involving events in the past of $\eventId$} (instead of hypothetical future branches), ensuring formulas have a definite runtime-computable truth value;
	\item Operators come in both quantified and unquantified forms, with the latter referring to the linear past on the same device \arevC{as the one of $\eventId$}, behaving similarly to past-LTL operators.
\end{itemize}
The modalities are inspired by terms like \emph{Yesterday}, \emph{Since}, \emph{Previously}, \emph{Historically}. We choose $\DX, \EX, \DU, \AU, \EU$ as primitive modalities, with the following informal meaning:
\begin{itemize}
	\item $\DX\formula$ means ``$\formula$ held in the \arevC{(immediately)} previous event \arevC{$\eprec(\eventId)$ on the same device $\devof(\eventId)$ as $\eventId$}'';
	\item $\EX\formula$ means ``$\formula$ held in \arevC{some neighbour event of $\eventId$}'';
	\item $\formula\DU\formulalt$ means ``$\formulalt$ held in some past event on \arevC{the same device $\devof(\eventId)$ as $\eventId$}, and $\formula$ has held \arevC{in events on $\devof(\eventId)$} since then'';
	\item similarly, $\formula\AU\formulalt$ (resp.~$\formula\EU\formulalt$) mean ``for all \arevC{(message)} paths (resp.~exists a path) \arevC{in $\aEventS$} reaching the current event, $\formulalt$ held in some event of the path and $\formula$ has held \arevC{on the path events} since then''.
\end{itemize}
The other operators can be derived from them through $\AX \formula \triangleq \neg \EX \neg \formula$; $\DF \formula \triangleq \top \DU \formula$ (similarly for $\AF$, $\EF$ with $\AU$, $\EU$); $\DG \formula \triangleq \neg \DF \neg \formula$ (similarly for $\AG$, $\EG$ with $\EF$, $\AF$).

\arevC{SLCS, on the other hand, is a modal logic where formulas are interpreted on nodes of a (possibly directed) graph, without a time component, thus predicating properties of space. When applied to a real-time distributed system, we can informally understand SLCS formulas as predicating properties of the \emph{current} state of the network, according to the knowledge that is available on each device (that may not be perfect). This is formalised by the notion of \emph{self-stabilisation} \cite{adsv:spatialrv}: monitors of SLCS formulas converge to the formula value in a finite time after the network and atomic propositions stabilise. Spatial modalities are divided into \emph{local} modalities (predicating properties of the neighbourhood of each device) and \emph{global} modalities (predicating properties of the whole network).}
The local modalities include:
\begin{itemize}
	\item $\SI \formula$ (interior): true at points where all neighbours satisfy $\formula$;
	\item $\SC \formula$ (closure): true at points where a neighbour satisfies $\formula$;
	\item $\SB$, $\SBI$ and $\SBC$ represent respectively \emph{boundary} (closure without interior), \emph{interior boundary} (set without the interior) and \emph{closure boundary} (closure without the set).
\end{itemize}
We take $\SC$ as the primitive modality and define the others through $\SI \formula \triangleq \neg \SC \neg \formula$, $\SB \formula \triangleq (\SC \formula) \wedge \neg (\SI \formula)$, $\SBI \formula \triangleq \formula \wedge \neg (\SI \formula )$, $\SBC \formula \triangleq (\SC \formula) \wedge \neg \formula$.
The global modalities are:
\begin{itemize}
	\item $\formula \SR \formulalt$ (reaches): true at the ending points of \arevC{(finite)} paths in the graph whose starting point satisfies $\formulalt$, and where $\formula$ holds on each point on the path;
	\item $\formula \ST \formulalt$ (touches): true at the end of paths whose start satisfies $\formulalt$ and where $\formula$ holds in the rest of the path;
	\item $\formula \SU \formulalt$ (surrounded by): true at points in an area satisfying $\formula$, whose closure boundary satisfies $\formulalt$;
	\item $\SG \formula, \SF \formula$ (everywhere, somewhere): true \arevC{in points such that} $\formula$ holds in every (resp.~some) point of every (resp.~some) incoming path \arevC{(thus, true in points such that $\formula$ holds everywhere/somewhere in the connected component of the point)}.
\end{itemize}
We take $\SR$ as primitive, and express the others by means of equivalences $\formula \ST \formulalt \triangleq \formula \SR \SC \formulalt$, $\formula \SU \formulalt \triangleq \formula \wedge \SI \neg (\neg \formulalt \SR \neg \formula)$, $\SF \formula \triangleq \top \SR \formula$, $\SG \formula \triangleq \neg \SF \neg \formula$.
}

Since information from space cannot be accessed directly, and is only available by exchanging and propagating data, the two logics \revB{described above} have very distinct features: while Past-CTL logic formulas are computed \emph{exactly} by their aggregate monitors, SLCS formulas are only computed \emph{approximately}. In particular, in \cite{adsv:spatialrv} the whole logic is implemented in \ac{xc} based on the modality $\SF$ (``somewhere''), which is implemented approximately through a distributed distance calculation.

\revB{In this scenario, we} compared the effectiveness of the predicates $\EP \textit{critic}$ (ever critic) and $\SF \textit{critic}$ (now critic SLCS) with a new approach in computing whether a Boolean value is somewhere true, by replicating Past-CTL operators. We thus wrote a generic algorithm replicator in FCPP:

\noindent\begin{minipage}{\textwidth}
\begin{lstlisting}[language=fcpp,aboveskip=0.35cm]
GEN(F) auto replicate(ARGS, F fun, size_t n, times_t t) { CODE
    size_t now = shared_clock(CALL) / t;
    auto res = spawn(CALL, [&](size_t i){
        return make_tuple(fun(), i > now - n);
    }, common::option<size_t>{now});
    for (auto x : res) if (x.first > now - n) now = min(now, x.first);
    return res[now];
}
\end{lstlisting}
\end{minipage}
\extLmcs{
This function creates a new replica of an algorithm \texttt{fun} every $t$ seconds, dropping older replicas so that no more than $n$ are present at each time. The result of the replication is the result of the oldest replica of the algorithm still alive. By carefully tuning the $t$ and $n$ parameters, we can ensure that the value returned is that of a replica that had just enough time to reach convergence, giving us the most recent view of what is the value of \texttt{fun} on recent values in the network. By replicating the $\EP$ operator from Past-CTL, we can thus turn \emph{``at some point in the past''} into another implementation of \emph{``somewhere right now''}:
}
\begin{lstlisting}[language=fcpp]
FUN bool somewhere(ARGS, bool f, size_t replicas, real_t diameter, real_t infospeed) { CODE
    return replicate(CALL, [&](){
        return logic::EP(CALL, f);
    }, replicas, diameter / infospeed / (replicas-1));
}
\end{lstlisting}
\extLmcs{
Given an estimate $D$ of the network diameter and of the speed $v$ at which information propagates, parameter $t$ is set to $\frac{D}{v (n-1)}$, which ensures that the oldest replica alive in a moment had just enough time to reach convergence.

\revD{
\begin{rem}[SLCS variants]
	In this paper, we follow SLCS as presented in \cite{adsv:spatialrv}. As already pointed out in \cite[Remark 2]{adsv:spatialrv}, that presentation is not exactly identical to the one given in \cite{DBLP:journals/corr/CianciaLLM16,DBLP:conf/ifipTCS/CianciaLLM14}. This is due to a different target application domain: abstract model checking for \cite{DBLP:journals/corr/CianciaLLM16,DBLP:conf/ifipTCS/CianciaLLM14}, and distributed monitoring for \cite{adsv:spatialrv}. The different target implies a different choice of the base model (continuous functions on closure spaces vs finite paths on directed graphs), although the two base models can be proven equivalent. Crucially, it also entails a different choice of primitive global operators (surrounded vs reachability), while tweaking the semantics of reachability in order to consider edges in the opposite direction, so that the properties can be computed through local interactions without a global oracle (that is not available in our model).
\end{rem}
}

\subsubsection{Setup: Scenario, Parameters, Metrics}

\revB{We consider a wide range of networks, varying the size in terms of (maximum) number of hops (parameter \emph{hops}) between $6$ and $16$, and varying the average number of devices per neighbourhood (parameter \emph{dens}) between $8$ and $18$, always with a connection range of $100 m$. Devices periodically compute the monitors about every second, with a standard deviation of the round time interval (parameter \emph{tvar}) ranging from $0\%$ to $40\%$, for $100$ seconds of simulated time. Devices move following randomly-generated checkpoints in their deployment area, with a \emph{speed} varying from $0$ to $20\%$ of the communication range per round.
Parameters \emph{tvar}, \emph{dens}, \emph{hops}, and \emph{speed} have the same meaning as explained in \Cref{subsec:msg-propagation-setup}.

In a first test, we set \emph{tvar}=$10$, \emph{dens}=$10$, \emph{hops}=$10$, and \emph{speed}=$10$, and measure the average of the devices with a \emph{true} value for four different predicates: \emph{critic}, \emph{ever critic} ($\EP \textit{critic}$), \emph{now critic SLCS} ($\SF \textit{critic}$), \emph{now critic replicated} ($\textrm{somewhere}(\textit{critic})$).
\revB{The number of replicas is set to $4$}.
We let the \emph{critic} predicate become true in the simulated times between $10$ and $30$ for a single device.

In four additional tests, we measure the percentage of time during executions (the average over all devices) when the computed values of $\SF \textit{critic}$ and $\textrm{somewhere}(\textit{critic})$ are wrong (with value $1$ representing a wrong value, and value $0$ a correct one). We let parameters \emph{tvar}, \emph{dens}, \emph{hops}, and \emph{speed} vary over ranges of relevant values.
}

\begin{figure*}[ht]
  \centering
  \includegraphics[width=0.5\figResizeFactor\linewidth{}]{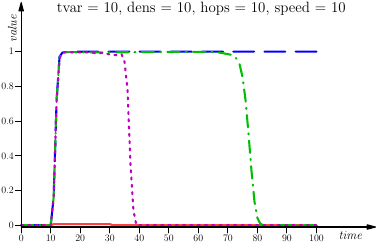}
  \includegraphics[width=\linewidth]{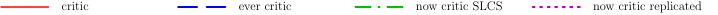}
  \caption{\extLmcs{Average truth value of spatio-temporal logic monitors over time.}} \label{fig:replicated:pastctl-time}
\end{figure*}

\subsubsection{Results}

\revB{
  Figure \ref{fig:replicated:pastctl-time} shows the value of the monitors over time. The data points are averaged over $400$ runs on networks with random initial positions of devices. All three monitors are very effective in rapidly propagating the information that a critic event is occurring throughout the whole network. Then, the \emph{ever critic} predicate never switches back to false by design. On the other hand, both approaches to computing \emph{now critic} successfully switch back to false some time after the criticality disappears at second $30$. The replicated approach, however, has a much faster reaction time, performing the switch around second $40$ while the SLCS formula only does it about $40$ seconds later, with a much slower reaction time.
}
\revB{
  As explained above, {\em ideally}, updating the \emph{now critic replicated} predicate when the \emph{critic} goes back to $\bot$ at second $30$ would just take the time to broadcast the new value to all the other nodes through multiple hops (i.e., local communications); in other words, it would take more or less the same amount of time as when the critic rises to $\top$ at time $20$. In practice, the approximate estimation of diameter $D$ and information speed $v$, and the limited number of replicas cause the observed delay.
  Instead, updating the \emph{now critic SLCS} (which uses a distributed distance computation under the hood) suffers from the well-known {\em rising value} problem (also known as {\em count-to-infinity}) \cite{audrito:coord:17}, which causes a significantly higher delay compared to \emph{now critic replicated}.
  This suggests that implementing the {\em somewhere} operator of SLCS exploiting \acp{dcp} would lead to an improved monitoring of SLCS formulas.
}

\begin{figure*}[ht]
  \centering
  \includegraphics[width=0.45\linewidth{}]{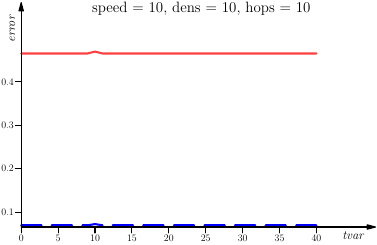}
  \includegraphics[width=0.45\linewidth{}]{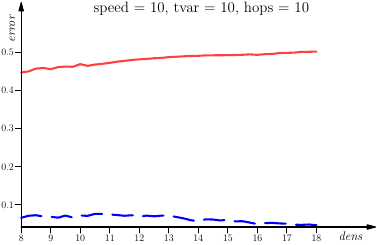}
  \includegraphics[width=0.45\linewidth{}]{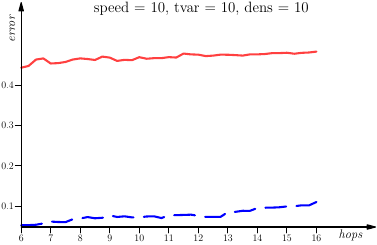}
  \includegraphics[width=0.45\linewidth{}]{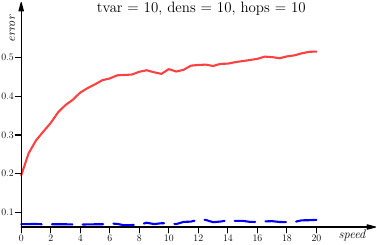}
  \includegraphics[width=0.45\linewidth{}]{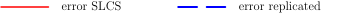}
  \caption{\extLmcs{\revB{Average error of spatio-temporal logic monitors, varying \emph{tvar} (top left), \emph{dens} (top right), \emph{hops} (bottom left), and \emph{speed} (bottom right).}}} \label{fig:replicated:pastctl}
\end{figure*}

\revB{
  Figure \ref{fig:replicated:pastctl} shows the percentage of erroneous values computed by monitors during executions, averaged over the network devices and the whole simulation time. Each data point is further averaged over $10$ runs on networks with different random seeds. It is easy to see that, for all four parameters \emph{tvar}, \emph{dens}, \emph{hops}, and \emph{speed}, and across the whole ranges in which we make them vary, the \emph{now critic SLCS} monitor is clearly significantly less accurate than the \emph{now critic replicated} one.
  This confirms that the results of the first test above hold in a very large number of different contexts.
}
}

\section{Discussion and Examples}\label{sec:disc}

In this section, we further discuss the proposed abstraction.
In particular, we delineate
 the characteristics of the proposed programming model for \acp{dcp} (\Cref{ssec:features}), and then provide examples of applications (\Cref{ssec:examples}) showing its versatility.


\subsection{Features of the abstraction and programming model}
\label{ssec:features}

\subsubsection{Progressive, live, and self-* construction of ensembles}
The \acp{dcp} have a \emph{dynamic} domain,
 that evolves progressively to include more or less devices.
The devices at the border of the \ac{dcp}
  can choose to expand it to (a subset of) their neighbours
  and the neighbours themselves can opt in or out.
Moreover, since evaluation of the program is repeated over time,
 the border is \emph{live},
 meaning that membership can be always re-evaluated,
 in order to consider the up-to-date context.
Conversely, members that are no longer interested in participating in the collective task, or that have completed the tasks associated to their role, can leave the process \revA{by returning $\falsevalue[]$ in the spawn routine, or even start process termination patterns as those investigated in \cite{DBLP:conf/acsos/AudritoCT22}.}

\subsubsection{Flexible regulation of collective process shape and state}
The shape and state of a \ac{dcp} can be regulated flexibly, by leveraging different kinds of mechanisms.
\extLmcs{
Two key mechanisms can be exploited to rule the participation of devices into \acp{dcp}:
 (i) the sharing of \acp{pid} with neighbours (i.e., the control of process propagation ``from the inside'', namely from the devices of the \emph{internal border} of the \ac{dcp}),
and (ii) the possibility of a device to \emph{quit} a process instance that has been received, preventing further propagation of it (i.e., the control of process shape ``from the outside'', namely from the devices of the \emph{external border} of the \ac{dcp}).

Moreover, }
 the state and shape of a process
 can be controlled at a collective level,
 as a result of a collective consensus or decision-making activity.
Also,
 the leader or owner of the \ac{dcp}
 may centralise some decision-making:
 e.g., it may gather statistics from its members \revA{(using adaptive collection algorithms \cite{DBLP:journals/cee/AudritoCDPV21})}, and use \revA{locally-computed} policies to decide whether to let more members join \revA{(sharing the local decision with a resilient broadcast algorithm \cite{vbdacp:survey})}.
Between fully centralised and fully decentralised settings,
  there are intermediate solutions based e.g. on a partitioning of the \ac{dcp} into sub-groups \revA{using partitioning mechanisms that can be applied at the aggregate programming level \cite{DBLP:journals/swarm/AguzziACDTV23,DBLP:conf/coordination/CasadeiPVN19}.}
The state can be used, for instance, to denote different \emph{phases} of a collective task~\cite{DBLP:conf/coordination/CasadeiVRA21}, hence it is important that all the members of the \ac{dcp} eventually become aware of the up-to-date situation.
Regarding shape control,
 further flexibility is provided by XC,
 thanks to differentiated messages to neighbours:
 this feature could be used to essentially control the direction of process propagation
 (e.g., by filtering, random selective choice, or any other ad-hoc mechanisms).
\extLmcs{
In future work,
 it would be interesting to explore how these mechanisms
 might support morphogenesis in modular robotics.
}

\subsubsection{Support for privacy-preserving collective computations}
The possibility in XC to send differentiated messages to neighbours (unlike classical field calculi~\cite{vbdacp:survey}),
 especially when supported infrastructurally through point-to-point communication channels,
 can also promote \emph{privacy} in collective computations.
\extLmcs{
By regulating the spread of processes ``from the inside'', devices that do not have to be aware of certain tasks
 are not exposed to such potentially sensible information.
Then,
 of course, a proper management of privacy (and security) would require addressing the issue
 throughout the software stack, i.e., also at the level of the middleware or platform sustaining the collective computation~\cite{DBLP:conf/goodtechs/PianiniCC0VO18}.
}

\subsection{Examples}
\label{ssec:examples}

Given its features, the \ac{dcp} abstraction could turn useful
 to program several kinds of higher-level distributed computing abstractions and tasks such as, for instance, the following.

\subsubsection{Modelling of teams or  ensembles of agents~\cite{DBLP:journals/taas/NicolaLPT14}}
A \ac{dcp} can represent, through its very domain, the set of agents that belong to a certain team or ensemble.
\extLmcs{
Consider:
}
\begin{lstlisting}[language=fcpp]
set<task_t> tasks = /* arbitrary task allocation */;
spawn(CALL, [&](task_t t){
  // add code solving the task
  // run the process where the task is allocated:
  nvalue<bool> nv_out = map_hood([&](device_t id){
    return t.contains(id);
  }, node.nbr_uid());
  return make_tuple(nullptr, nv_out);
}, tasks);
\end{lstlisting}
\extLmcs{
Here, there will be a different process instance for each different task, and all the devices assigned to the same task will form a team. Notice, however, that two devices of the team would be able to communicate within the process only if they are connected by a path of devices also belonging to the same team.
The \ac{dcp} associated to a team
} can spread around the system to gather and (re-)evaluate a membership condition, to effectively recruit agents into different organisational structures~\cite{DBLP:journals/ker/HorlingL04}.
The former mechanism
 is directly supported by our XC implementation,
 through the notion of \emph{differentiated messages} to neighbours, enabled by \excvalues{}.
Concurrent participation to multiple teams
 is directly supported by the fact that
 a single device can participate in an arbitrary number of \acp{dcp}. \revA{As participation to multiple \acp{dcp} leads to increased resource requirements (in both computation time and message size), the programmer has to take into account performance issues when designing the generation and propagation logic of concurrent \acp{dcp}. However, the fully asynchronous and resilient nature of XC implies that some additional slack can be used on top of resource bounds posed by the architecture, as longer round execution or message exchange time (or even a device crash) can be handled seamlessly by the XC programming model.}
Last but not least,
 the activity within a \ac{dcp} can be used to support the coordination \emph{within} the ensemble it represents, e.g., through gossip or  information spreading algorithms, whose scope is limited to the domain of the \ac{dcp}; therefore, it may be useful also for \emph{privacy-preserving} computations.

\begin{figure}[t]
\newcounter{nn}\setcounter{nn}{0}
\pgfmathtruncatemacro{\nv}{\value{nn}}
\newcommand{\pointIntersection}[4]{
\path [name intersections={of=#2 and #3, by=#4}]
}

\newcommand{\cone}[8]{
  \stepcounter{nn}
  \node[point] (#8src) [#2,label={#1}] {};
  \draw[name path=#8,fill=#7!40,opacity=0.4,draw=none] (#8src.center) -- ($(#8src)+(#3cm,#4cm)$) -- ($(#8src)+(#5cm,#4cm)$) -- ($(#8src)+(#6cm,0cm)$) -- ($(#8src)+(#5cm,-#4cm)$) -- ($(#8src)+(#3cm,-#4cm)$) -- cycle;
}
\newcommand{\conew}[9]{
  \stepcounter{nn}
  \node[point] (#8src) [#2,label={#1}] {};
  \draw[name path=#8,fill=#7!40,opacity=0.4,draw=none,#9] (#8src.center) -- ($(#8src)+(#3cm,#4cm)$) -- ($(#8src)+(#5cm,#4cm)$) -- ($(#8src)+(#6cm,0cm)$) -- ($(#8src)+(#5cm,-#4cm)$) -- ($(#8src)+(#3cm,-#4cm)$) -- cycle;
}
\newcommand{\conefullw}[9]{
  \stepcounter{nn}
  \node[point] (#9src) [#1] {};
  \draw[name path=#9,opacity=0.4,draw=none,pattern=north east lines,pattern color=black] (#9src.center) -- ($(#9src)+(#2cm,#3cm)$) -- ($(#9src)+(#4cm,#3cm)$) -- ($(#9src)+(#5cm,#6cm)$) -- ($(#9src)+(#4cm,-#7cm)$) -- ($(#9src)+(#8cm,-#7cm)$) -- cycle;
}
\newcommand{\fcone}[8]{
  \stepcounter{nn}
  \node[point] (#8src) [#2,label={#1}] {};
  \draw[name path=#8,fill=#7!40,opacity=0.4,draw=none,frastagliato] (#8src.center) -- ($(#8src)+(#3cm,#4cm)$) -- ($(#8src)+(#5cm,#4cm)$) -- ($(#8src)+(#6cm,0cm)$) -- ($(#8src)+(#5cm,-#4cm)$) -- ($(#8src)+(#3cm,-#4cm)$) -- cycle;
}
\newcommand{\fconestart}[6]{
  \stepcounter{nn}
  \node[point] (#6src) [#2,label={#1}] {};
  \draw[name path=sa#6,fill=#5!40,opacity=0.4,draw=none,frastagliato] (#6src.center) -- ($(#6src)+(#3cm,#4cm)$);
  \draw[name path=sb#6,fill=#5!40,opacity=0.4,draw=none,frastagliato] (#6src.center) -- ($(#6src)+(#3cm,-#4cm)$);
}
\newcommand{\fpcone}[8]{
  \stepcounter{nn}
  \node[point] (#8src) [label={#1}] at (#2) {};
  \draw[name path=#8,fill=#7!40,opacity=0.4,draw=none,frastagliato] (#8src.center) -- ($(#8src)+(#3cm,#4cm)$) -- ($(#8src)+(#5cm,#4cm)$) -- ($(#8src)+(#6cm,0cm)$) -- ($(#8src)+(#5cm,-#4cm)$) -- ($(#8src)+(#3cm,-#4cm)$) -- cycle;
}
\newcommand{\fpconez}[8]{
  \stepcounter{nn}
  \node[point] (#8src) [#1] at (#2) {};
  \draw[name path=#8,fill=#7!40,opacity=0.4,draw=none,frastagliato] (#8src.center) -- ($(#8src)+(#3cm,#4cm)$) -- ($(#8src)+(#5cm,#4cm)$) -- ($(#8src)+(#6cm,0cm)$) -- ($(#8src)+(#5cm,-#4cm)$) -- ($(#8src)+(#3cm,-#4cm)$) -- cycle;
}
\tikzset{-,
  point/.style={circle,fill=black, inner sep=1.8pt, minimum size=2pt, outer sep=0.5mm},
  event/.style={circle,fill=red, inner sep=3pt, minimum size=2pt, outer sep=0.5mm},
  past/.style={event, fill=ddarkgreen},
  present/.style={event,fill=gray},
  future/.style={event,fill=blue},
  locality/.style={circle,fill=white, inner sep=2pt, minimum size=2pt, outer sep=0.5mm},
  locality1/.style={locality,fill=orange!50},
  locality2/.style={locality,fill=gray!50},
  locality3/.style={locality,fill=blue!40},
  locality4/.style={locality,fill=black!30!green},
  locality5/.style={locality,fill=purple!50},
  d/.style={circle,fill=gray, inner sep=2pt, minimum size=2pt},
  da/.style={circle,draw,label={[left]$1$}},
  db/.style={circle,draw,label={[left]$2$}},
  dc/.style={circle,draw,label={[left]$3$}},
  lbl/.style={font=\footnotesize},
  state/.style={circle,draw},
  statein/.style={circle,draw,line width={2pt}},
  stateout/.style={circle,draw,line width={1pt},fill=black!16},
  leadstov/.style={draw,decorate,decoration={snake, amplitude=1.95mm, post=lineto, post length=2mm, segment length=1cm, pre length=0.4cm},->,>=stealth'},
  leadsto/.style={draw,decorate,decoration={snake, amplitude=0.35mm, post=lineto, post length=2mm, segment length=1.3mm},->,>=stealth'}, 
  leadstopast/.style={leadsto,draw=ddarkgreen},
  leadstopresent/.style={leadsto,draw=black},
  leadstofuture/.style={leadsto,draw=blue},
  frastagliato/.style={draw,decorate,decoration={snake, amplitude=0.45mm, post=lineto, post length=0mm, segment length=2.3mm},->,>=stealth'},
}
\begin{center}
	\def\tlen{6.6cm}
	\def\slen{7cm}
	\def\ta{0.1*\tlen}
	\def\sa{0.1*\tlen}
	\scalebox{0.8}{
	\begin{tikzpicture}[node distance=0.8cm and 0.8cm] 
		\node[] (0) [] {};
		\node[] (1) [right=1.3*\tlen of 0] {};
		\node[] (2) [above=0.99*\slen of 0] {};

		\draw[->] (0) -- (1) node[below, midway, xshift=-1cm] {time};
		\draw[<->] (0) -- (2) node[above, midway, rotate=90] {space / devices};

		\cone{$\epsilon_{out}$}{above right = 2cm and 1.1cm of 0}{1.5}{1.5}{6.6}{6.6}{red}{cone1}
		\cone{$\epsilon_{in}$}{above right = 4.5cm and 0.2cm of 0}{1}{2}{6.6}{6.6}{blue}{cone2}
		\pointIntersection{$\epsilon_{M}$}{cone1}{cone2}{m1};
		\fcone{$\epsilon_{M}$}{above right= 0cm and 0cm of m1}{1.1}{1.1}{1.8}{1.8}{black}{cone3}
		\cone{$\epsilon_{in}'$}{above right = 0.4cm and 1.5cm of 0}{1}{0.6}{7}{7}{blue}{cone4}
		\pointIntersection{$\epsilon_{M2}$}{cone4}{cone1}{m2};
		\fcone{$\epsilon_{M}'$}{above right= -0.1cm and -0.1cm of m2}{1.1}{0.9}{1.2}{1.2}{black}{cone5}
		\fcone{$\epsilon_{C}$}{above right = 2cm and 3.5cm of 0}{1.5}{3.0}{2}{2}{red}{cone6}
		\draw[->,leadstopresent,gray] (cone2src) to [bend right=8] (cone3src);
		\draw[-,dashed] (cone1src) -- (cone6src);
		\path[name path=vrecv] (cone2src) -- ($(cone2src)+(5cm,0cm)$);
		\pointIntersection{}{cone6}{vrecv}{vrinc};
		\draw[-,dashed] (cone2src) -- (vrinc);
		\fpconez{label={[xshift=-0.2cm]$\epsilon^\dagger_{in}$}}{vrinc}{1.5}{3.0}{2}{2}{blue}{cone7}
		\path[name path=vack] (cone6src) -- ($(cone6src)+(3cm,0cm)$);
		\pointIntersection{}{cone7}{vack}{vackp};
		\draw[-,dashed] (cone6src) -- (vackp);
		\fpconez{label={[yshift=-0.7cm,xshift=-0.24cm]$\epsilon^\dagger_{out}$}}{vackp}{1.2}{3}{1.5}{1.5}{red}{cone8}
		\draw[->,leadstopresent,gray] (cone3src) -- (cone6src);
		\draw[->,leadstopresent,gray] (cone6src) to [bend right] (cone7src);
		\draw[->,leadstopresent,gray] (cone7src) to [bend left] (cone8src);
		\path[name path=vtop] (cone7src) -- ($(cone7src)+(3.2cm,0cm)$);
		\pointIntersection{}{cone8}{vtop}{vtopp};
	\end{tikzpicture}
	}
\end{center}
\caption{Graphics of interacting \acp{dcp} modelling \revB{spatiotemporal} tuple operations.
Each \ac{dcp} is denoted as a trapezoid-like shape that springs out at a certain event (a round by a single device).
Notation: $\epsilon_{in}$ and $\epsilon_{out}$ mean that the event generates a process modelling an \texttt{in} (retrieval) and \texttt{out} (writing) tuple operation, respectively; $\epsilon_{M}$ means that a matching \texttt{out} tuple for an \texttt{in} operation has been found; $\epsilon_{C}$ means that the \texttt{out} tuple process has reached consensus about the \texttt{in} process to serve; the $\dagger$ superscript denotes a termination event, starting a process to close an existing process.
}
\label{graphics-in}
\end{figure}

\subsubsection{Space-based coordination (e.g., spatiotemporal tuples~\cite{DBLP:conf/coordination/CasadeiVRA21})}
A \ac{dcp} could also be attached to a spatial location---to implemented \emph{spatially-attached processes}.
\extLmcs{
Consider an application that needs to collectively monitor some environmental quantities in
 a set of sensible regions of a smart city.
This could be encoded as follows:
} 
\pagebreak
\begin{lstlisting}[language=fcpp]
std::vector<region> sensible_regions = { region(...), region(...), ... };
std::vector<region> matching_regions;

for (region const& r : sensible_regions)
	if (nearby(CALL, r)) matching_regions.push_back(r);

spawn(CALL, /* process logic */, matching_regions);
\end{lstlisting}

These ideas could be leveraged to support space-based coordination, or to implement coordination models like \emph{spatiotemporal tuples}~\cite{DBLP:conf/coordination/CasadeiVRA21},
 whereby tuples and tuple operations can be emitted to reside at
 or query a particular spatial location.
To implement the spatiotemporal tuples model,
 an \ac{dcp} instance can be used to
 represent a single
 \texttt{out} (writing),
 \texttt{rd} (reading),
 and
 \texttt{in} (retrieval)
 operation---see \Cref{graphics-in} for a visual example.
A tuple is denoted by its \texttt{out} process: it exists as long as its \ac{dcp} is alive in some device.
Creating tuples that reside at a fixed spatial location/area (e.g., as described by geodetic coordinates)
 or that remain attached to a particular mobile device is straightforward.
In the former case,
 the \ac{dcp} membership condition is just that the device's current location is inside
 or close by the provided spatial location.
In the latter case,
 the \ac{dcp} membership condition is just that
 the device's current distance to the \ac{dcp} source device (which may be computed by a simple gradient) is within a certain threshold.
We may call these \emph{node-attached processes}: as a node moves, a \ac{dcp} attached to it can follow through, to support collective contextual services; for instance, a node may recruit other nodes and resources for mobile tasks.

\subsubsection{Creation of adaptive system structures to support communication and coordination}\label{ssec:scr}
The ability of \acp{dcp} to capture both the formation evolution and the collective activity
 of a group of devices within a pervasive computing system
 can be leveraged to create resilient structures
 supporting non-local or system-wide coordination.
For instance,
 this can be used to implement messaging channels in a peer-to-peer network of situated neighbour-interacting devices (e.g., in a smart city)~\cite{casadei2019aggregate}:
 the channel consists only of the devices between the source and the destination of a message,
 hence avoiding expensive gossip or flooding processes that would (i) consume resources of possibly all the devices in the system,
 and (ii) exacerbate privacy and security issues.
\extLmcs{
This idea could be implemented as follows:
}
\begin{lstlisting}[language=fcpp]
bool source = // ...
bool destination = // ...
real_t width = // ....
bool in_channel = channel(CALL, source, destination, width);
option<device_t> generation_pids;
if (in_channel) generation_pids = broadcast(CALL, source, node.uid);
spawn(CALL, /* process logic */, generation_pids);
\end{lstlisting}

\begin{figure}
\centering
\includegraphics[width=0.69\textwidth]{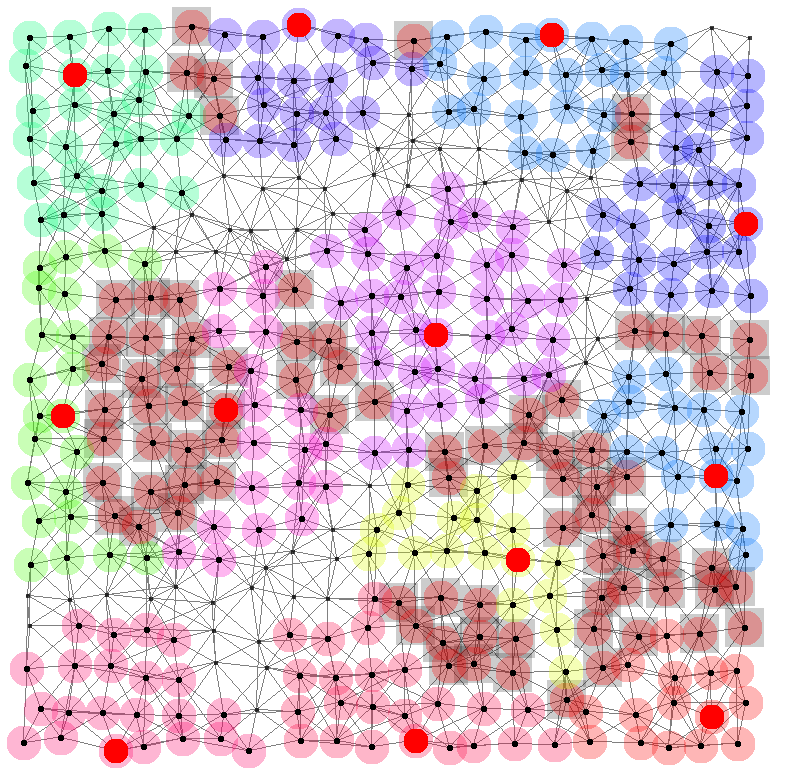}
\caption{\extLmcs{SCR pattern: snapshot from a simulation. Notation: black dots and links represent the network of devices; large red circles denote leaders; semi-transparent coloured circles denote running processes (different colours for different instances) as well as regions; the square semi-transparent shadows are used to highlight devices that run two or more process instances, hence denoting the overlapping of processes.}}
\label{fig:scr}
\end{figure}

As another example, consider the \emph{Self-organising Coordination Regions (SCR)} pattern~\cite{DBLP:conf/coordination/CasadeiPVN19}:
 it is a mechanism to control the level of decentralisation in systems to flexibly support situated tasks, based on
  (i) decentralised leader election~\cite{DBLP:conf/acsos/PianiniCV22};
  (ii) creation of areas around the leaders to basically partition the overall system into manageable regions~\cite{DBLP:conf/ceemas/WeynsH03};
  and
  (iii) supporting intra-region and inter-region coordination e.g. by means of \emph{information flows}~\cite{DBLP:conf/saso/WolfH07}.
Note that traditional solutions based on field calculi~\cite{DBLP:conf/coordination/CasadeiPVN19}
 do not easily allow for the partitions to \emph{overlap}:
 this, instead, could be desired for fault-tolerance, flexibility, and improved interaction between adjacent regions,
 and it turns out to be easily implementable using \acp{dcp}.
\extLmcs{
An implementation of SCR with overlapping regions could be coded as follows:
}
\makeatletter
\newcommand\currentStyle@lstparam{}
\lst@AddToHook{Output}{\global\let\currentStyle@lstparam\lst@thestyle}
\lst@AddToHook{OutputOther}{\global\let\currentStyle@lstparam\lst@thestyle}
\makeatother
\makeatletter
\newcommand{\crefe}[1]{\currentStyle@lstparam\autoref{#1}} 
\makeatother
\begin{lstlisting}[mathescape,language=fcpp]
real_t mean_area_size = // ...
bool leader = leader_election(CALL, mean_area_size);
// Now each leader starts a concurrent process that may overlap with others
common::option<device_t> gen;
if (leader) gen = node.uid;
spawn([&](device_t id, bool is_leader, real_t region_size){
  bool region_leader = leader == node.uid and is_leader;
  real_t g = gradient(CALL, region_leader); // cf. $\crefe{ex:gradient}$
  (region_leader || g < region_size, distributed_sensing(CALL, g))
}, gen, leader, 100.0);
\end{lstlisting}
\extLmcs{
In this case, a leader can also act as a regular member of another leaders' region.
\Cref{fig:scr} shows a snapshot from a simulation of this SCR pattern.
}

\extLmcs{

\section{Related Work}\label{sec:rw}

This paper presents the \emph{\acl{dcp} (\acs{dcp})} abstraction and its implementation in XC.
%
%
In the literature, several macro-level or collective-level abstractions~\cite{Casadei2023} have been proposed
 to capture or promote the emergence of coordinated behaviour of whole groups of interacting agents.
%
%

\paragraph{\emph{Topological and spatial abstractions}.}
These abstractions exploit space as a way to drive collective behaviour.
For instance, in \emph{SpatialViews}~\cite{DBLP:conf/pldi/NiKSI05},
 groups of nodes (i.e. ensembles) can be matched
 against spatiotemporal expressions
 (which define the scope and resolution for selection)
 and iterated upon to access or command them.
In \emph{Abstract Regions}~\cite{DBLP:conf/nsdi/WelshM04}, regions work as collective communication interfaces.
In XC it is also possible to express formulas for matching spatiotemporal regions (e.g., by querying device sensors) and hence the devices therein.
In particular, the domain of a \ac{dcp} fully denotes a spatiotemporal region, with the process logic embedding the rules for setting up, evolving, and tearing down the region; several kinds of shapes and dynamics can be programmed (cf. \Cref{sec:disc}).
Languages tailored to shape and pattern formation
 in systems of neighbour-iteracting devices
 have also been proposed,
 such as \emph{Origami Shape Language (OSL)}~\cite{DBLP:phd/ndltd/Nagpal01},
 where shapes are constructively described as a sequence of straight folds performed on a square sheet;
 however, these approaches focus on shape creation and are not general-purpose.

\paragraph{\emph{Ensembles}~\cite{DBLP:journals/taas/NicolaLPT14,DBLP:conf/cbse/BuresGHKKP13}} Ensembles are dynamic composites of devices. Ensembles can be explicitly or implicitly defined (e.g., dynamically as result of applying attribute-based formation rules),
 and can be used to structure and refer groups of coordinated entities (e.g., in terms of roles) or to structure coordination itself, for instance by denoting them as targets of communication.
In \emph{Distributed Emergent \revB{Ensembles} of Components (DEECo)}~\cite{DBLP:conf/cbse/BuresGHKKP13}, ensembles are formed according to \emph{membership conditions}.
In \emph{Service Component Ensemble Language (SCEL)}~\cite{DBLP:journals/taas/NicolaLPT14} and \emph{Attribute-based Communication (AbC)}~\cite{abd2020programming-cas-attribute-based},
 ensembles are specified according to predicates on the attributes that the components expose.
On the other hand,
 in Bucchiarone et al.~\cite{DBLP:conf/birthday/BucchiaroneM19},
 ensembles are groups with roles to be filled,
 and ensemble configurations are described as hypergraphs
 that specify cells (i.e., roles to be taken by participants), adaptors (i.e., coordination rules), and fragments (provided functionality).
There are also practical languages for swarm robotics
 where ensemble-like abstractions are exposed.
In Buzz~\cite{DBLP:journals/software/PinciroliB16},
 swarms are first-class abstractions
 that can be handled by union, intersection, iteration, etc.
In Voltron~\cite{Mottola2014voltron}, a language for drone sensor networks,
 there is a notion of a \emph{team} abstraction
 that can be the target of action commands
 and that automatically handles the underlying drones according to the (spatiotemporal) tasks at hand.
Most of similar approaches focus on task orchestration, leveraging centralised coordinators.
For instance, Dolphin~\cite{lima2018dolphin} has a notion of vehicle set which is similar to a swarm in Buzz,
 whereas TeCoLa~\cite{Koutsoubelias2016tecola}, a framework for coordination of robot teams, provides a notion of a ``mission group'' that is automatically managed according to the sets of services supported by the robots.

\paragraph{\emph{Collective communication interfaces}.}
Collective communication interfaces are abstractions for sending data to or receiving data from groups of devices.
These abstractions usually enable to  flexibly express the targets of communications actions, e.g., via attributes~\cite{DBLP:journals/taas/NicolaLPT14,abd2020programming-cas-attribute-based}.
Indeed, this abstraction is very much related to the discussed notion of ensemble, as an ensemble can work as a collective communication endpoint.

\paragraph{\emph{Collective-based tasks}.}
Collective-based tasks are abstractions keeping track of the lifecycle and state of tasks assigned to whole collectives.
These are generally used in \emph{orchestration} approaches.
An example is the \emph{SmartSociety} platform~\cite{DBLP:journals/tetc/ScekicSVRTMD20}, which is aimed at heterogeneous collectives involving both humans and machines.
In XC, it is also possible to denote collective tasks as regular values, and use gossip-like library functions to distribute them over arbitrary target domains.

\paragraph{\emph{Collective data structures.}}
Collective data structures associate data to the members of a collective.
A major example is the notion of a \emph{computational field}~\cite{Mamei:2004a,vbdacp:survey}: statically, it is a map from devices to values;
 by a dynamic perspective, it is a map from events (i.e., devices across space-time) to values.
While the long-standing related notion of \emph{artificial potential fields}~\cite{DBLP:conf/icra/Warren89} associates values with the environment, computational fields associates values with devices.
However, it can be noted that, since devices may cover or fill an environment, these two notions may overlap, even though with computational fields there is the idea that devices collaborate to provide local values for these fields to emerge.
Computational fields can be used to represent collective inputs/outputs, as well as intermediate values for larger computations; this insight leads to the following notion.

\paragraph{\emph{Aggregate computations.}}
Collective computations are the computations performed by collectives of computational entities.
Often, these involve interaction among such entities.
One major example of collective computations are \emph{aggregate computations}~\cite{vbdacp:survey}, which can be implemented as \emph{field computations} in \emph{field calculi}~\cite{vbdacp:survey}; in this case, programs are represented as functions mapping input computational fields to output computational fields, while implicitly handling coordination.
XC provides a generalisation of field calculi in terms of a single construct, \lstinline[language=xc]|exchange|, and \excvalues{}. 
The main benefit of these approaches lies in the possibility of  representing whole collective behaviours as reusable functions and using functional composition to effectively compose larger collective behaviours from smaller ones.
%

\paragraph{\emph{Aggregate processes.}}
Aggregate processes are dynamic \emph{aggregate computations}~\cite{vbdacp:survey} on evolving domains of devices, formalised within the \ac{fc}~\cite{DBLP:conf/acsos/AudritoCT21a,DBLP:conf/acsos/AudritoCT22,casadei2019aggregate,
casadei2021engineering}.
The \ac{dcp} abstraction presented in this work (cf. \Cref{sec:disc-abs})
 is essentially a generalisation of aggregate processes for the XC.
The implementation in XC is also more expressive,
  for it enables to \emph{control how processes spread from a node to neighbours}, rather than controlling shape evolution based on \emph{opt-out} (i.e., the receiving device decides to quit the process) which is the only option in aggregate processes.
}

\section{Conclusion and Future Work}\label{sec:conc}

\extLmcs{
In this article,
 we have described the \emph{\acl{dcp} (\acs{dcp})} abstraction,
 which aims to structure dynamic and concurrent collective activities
  on ensembles of neighbour-interacting devices running asynchronously in sense--compute--interact rounds.
In particular, \acp{dcp}
 model decentralised collective tasks
 that can spread, shrink, and move
 across the devices of the collective system
 that sustain their execution.
We have presented the abstraction
 in the general framework of event structures
 and provided a formalisation in the \emph{eXchange Calculus (XC)}.
XC is a minimal, functional core language
 for self-organising systems
 based on \emph{Neighbouring Values},
 which supports fine-grained management of \acp{dcp}
 by precisely controlling information sharing within a device's neighbourhood.
The implementation consists of a construct $\spawnxc$
 that handles the generation and execution of \ac{dcp} instances.
The construct has been included in the FCPP language
 and the mechanism has been studied using two case studies of message propagation and distributed monitoring,
  showing its functionality and versatility.
A discussion and examples have also been provided
 to convey how \acp{dcp} can be programmed
 to support various kinds of self-organisation patterns.

The proposed abstraction and implementation
 can be further developed along different directions.
On one hand,
 \acp{dcp} could be more deeply analysed by a dynamical perspective, e.g. by studying the conditions for expansion and termination across different instantiations of the execution model.
Then, its suitability for peculiar distributed computing scenarios such as epidemic modelling and morphogenesis could be investigated.
On a more practical side,
 common patterns of \acp{dcp} usage
 could be encapsulated into a library of reusable functions,
 to foster application development.
}

%
%
%

\section*{Acknowledgment}

This publication is part of project NODES, which has received funding from the MUR -- M4C2 1.5 of PNRR with grant agreement no. ECS00000036. The work was also supported by the Italian PRIN project ``CommonWears'' (2020HCWWLP).

\bibliographystyle{alphaurl}
\bibliography{biblio}

\end{document}